\documentclass[preprint,12pt]{elsarticle}

\usepackage{amsmath}

\usepackage{amssymb}
\usepackage{boondox-cal}
\usepackage{graphicx} 
\usepackage{color}
\usepackage{float}
\usepackage{enumerate}
\usepackage{indentfirst}

\usepackage{indentfirst}
\usepackage{booktabs}
\usepackage{diagbox}
\usepackage{braket}
\usepackage{subfigure}
\usepackage{appendix}

\usepackage{lineno}
\journal{}

\begin{document}

\begin{frontmatter}

\title{The robustness of skyrmion numbers of structured optical fields in atmospheric turbulence}

\author[inst1,inst3]{Liwen Wang}

\affiliation[inst1]{organization={Key Laboratory of Quantum Information},
            addressline={University of Science and Technology of China}, 
            city={Hefei},
            postcode={230026}, 
            state={Anhui},
            country={China}}

\author[inst1,inst3]{Sheng Liu\corref{corresponding}}
\ead{shengliu@ustc.edu.cn}
\author[inst1,inst3,inst4]{Geng Chen}
\author[inst1,inst2,inst3,inst4]{Yongsheng Zhang\corref{corresponding}}
\ead{yshzhang@ustc.edu.cn}
\author[inst1,inst3,inst4]{Chuanfeng Li}
\author[inst1,inst3,inst4]{Guangcan Guo}

\cortext[corresponding]{Corresponding authors.}

\affiliation[inst2]{organization={Institute of Artificial Intelligence},
            addressline={Hefei Comprehensive National Science Center}, 
            city={Hefei},
            postcode={230088}, 
            state={Anhui},
            country={China}}

\affiliation[inst3]{organization={CAS Center for Excellence in Quantum Information and Quantum Physics},
            city={Hefei},
            postcode={230026}, 
            state={Anhui},
            country={China}}

\affiliation[inst4]{organization={Hefei National laboratory},
            addressline={University of Science and Technology of China}, 
            city={Hefei},
            postcode={230088}, 
            state={Anhui},
            country={China}}

\begin{abstract}
The development of vector optical fields has brought forth numerous applications. Among these optical fields, a particular class of vector vortex beams has emerged, leading to the emergence of intriguing optical skyrmion fields characterized by skyrmion numbers. The optical skyrmion fields are well-defined by their effective magnetization and possess topologically protected configurations. It is anticipated that this type of optical structure can be exploited for encoding information in optical communication, even under perturbations such as turbulent air, optical fibers, and even general random media. In this study, we numerically demonstrate that the skyrmion numbers of optical skyrmion fields exhibit a certain degree of robustness to atmospheric turbulence, even though their intensity, phase and polarization patterns are distorted. Intriguingly, it is also observed that a larger difference between the absolute values of two azimuthal indices of the vectorial structured light field can lead to a superior level of resilience. These properties not only enhance the versatility of skyrmion fields and their numbers, but also open up new possibilities for their use in various applications across noisy channels.
\end{abstract}

\begin{graphicalabstract}
\centering
\includegraphics[width=6.5in]{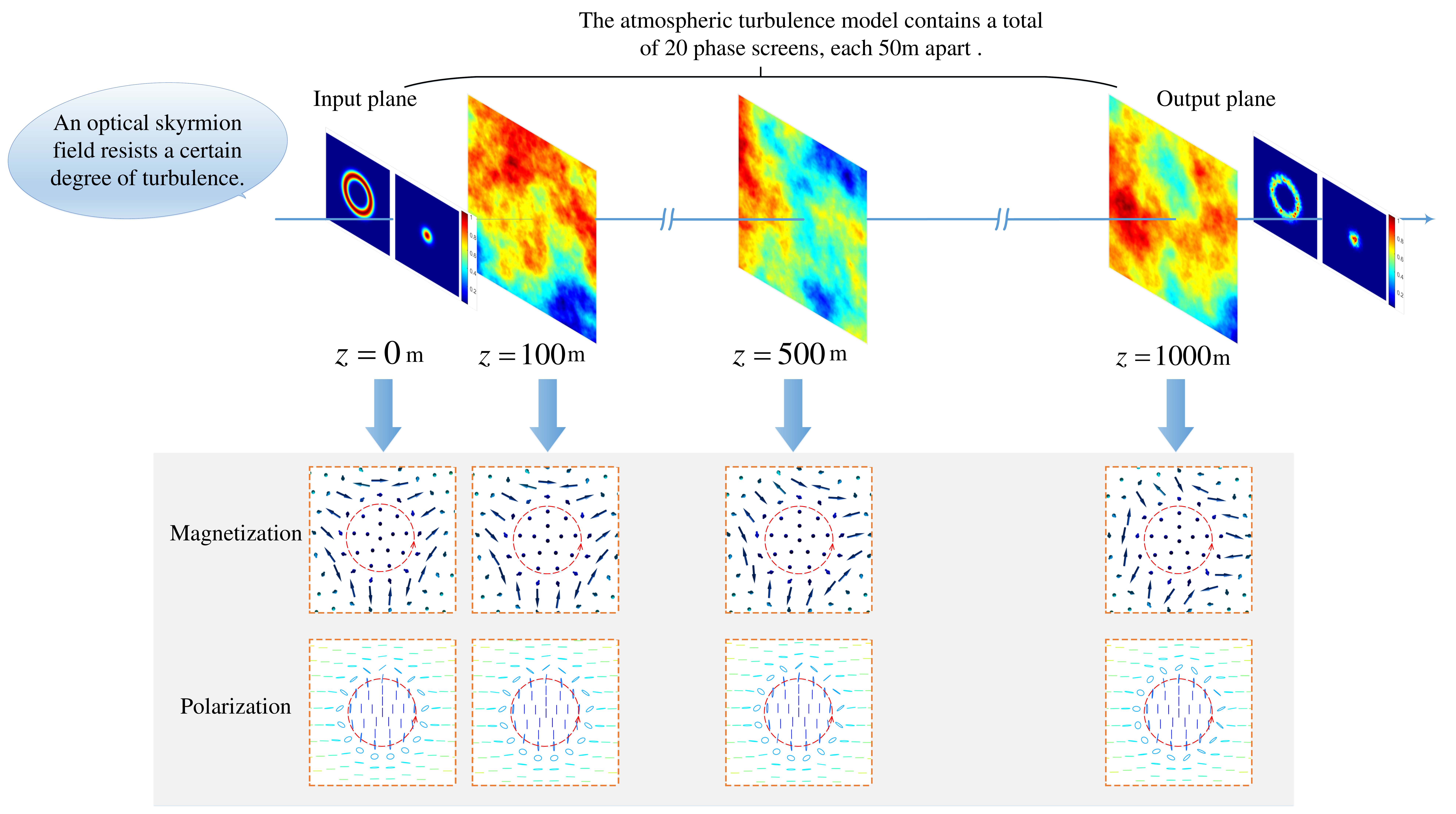}
\end{graphicalabstract}

\begin{highlights}
\item Skyrmions have topologically protected textures, providing an application prospect.
\item Skyrmion numbers are topological quantities and rather robust against air turbulence.
\item Different combinations of two spatial modes result in different levels of robustness.
\end{highlights}


\end{frontmatter}

\section{Introduction}
\noindent The transverse spatial modes of light provide a great resource and an additional degree of freedom for free-space optical (FSO) communication and information encoding \cite{krenn2016twisted,gibson2004free,willner2015optical,willner2021orbital}. In general, we can consider this type of spatial modes carrying orbital angular momentum (OAM) as the information carrier and such modes are referred to OAM modes that take on spiral wavefront and phase singularity \cite{padgett2017orbital}. OAM modes are a set of distinct, quantized states of angular momentum associated with the spatial distribution of light waves and the phase of such beams is $\exp(il\phi)$, where $l$ represents the OAM quantum number and is also known as the topological charge number. Here, $l$ can be any integer value \cite{allen1992orbital,chen2015research}. Thus, these modes theoretically have infinite dimensions that can be applied to high-dimensional systems and improve the channel capacity \cite{krenn2016twisted,wang2012terabit,krenn2014generation,huang2014100,malik2016multi,ren2016experimental,sit2017high,cozzolino2019orbital,liu2020multidimensional}. There are many ways to generate desired OAM beams such as photonic integrated devices \cite{cai2012integrated,zhou2018generating,xiao2016generation}, spatial light modulators (SLMs) \cite{lin2006synthesis,bolduc2013exact}, spiral phase plates (SPPs) \cite{beijersbergen1994helical,rafighdoost2017spirally}, metamaterials and metasurfaces \cite{xiaodong2021review,zhao2018generating}, mode convertors \cite{beijersbergen1993astigmatic}, Q-plates \cite{marrucci2006optical,karimi2009efficient} etc. With OAM-based communication links, FSO transmission has been developed in a spurt. In 2014, the total transmission rate through the channel has been increased to 100Tbit/s by Huang \textit{et al.} \cite{huang2014100} and Wang's team reached a total rate of 1.036Pbit/s with more OAM multiplexing in the same year \cite{wang2014n}. Consequently, using of OAM modes has a great contribution to the field of FSO communication.\par 
However, atmospheric turbulence is an important factor affecting optical communication \cite{anguita2007spatial,anguita2008turbulence,ren2013atmospheric,doster2016laguerre}. Atmospheric turbulence originates from the variations in temperature and pressure resulting in the refractive index fluctuations in air \cite{andrews2005laser}. In 1941, Kolmogorov introduced a statistical turbulence theory to interpret the stochastic behavior of atmospheric turbulence, so the well-defined Kolmogorov power-law spectrum is widely used in theoretical calculations that describe phase fluctuations caused by the variations in the refractive index of atmosphere for optical fields \cite{kolmogorov1941degeneration,kolmogorov1991dissipation}. We exploit the modified spectrum proposed by Andrew based on the Kolmogorov theory to generate thin random phase screens and employ the split-step method with subharmonics to simulate beam propagation through the atmosphere turbulence under realistic terrestrial conditions \cite{schmidt2010numerical}. Obviously, in turbulent air, cross talk among different OAM modes is inevitable, and their amplitude, phase and polarization patterns are disrupted due to the presence of atmospheric turbulence. So the channel capacity will decrease and optical communication will also encounter certain disturbances. To address this issue, one approach is to identify a particular light field resistant to external disturbances.\par
Recently, there has been growing interests in utilizing vector optical beams for FSO communication, information encoding and remote sensing \cite{rosales2018review,zhu2021compensation,nape2022revealing,singh2023robust,cheng2009propagation}. And some researchers have used `vectorness' of vector beams as an information carrier. A type of vector vortex beams, known as skyrmionic beams, which combine spatial and polarization dimensions, are believed to be a promising option against disturbances because of topological features from spatial variations in amplitude and polarization \cite{gao2020paraxial,zhu2021synthesis}. Their topological properties are typically characterized by corresponding skyrmion numbers, which can be directly computed using the expected values of Pauli operators. Skyrmions were originally used in magnetic systems \cite{fert2017magnetic}. Magnetic skyrmions are a class of topological stable textures and are expected to be robust for local perturbations. Later researchers have found that an optical skyrmion can be created by some ways and such a skyrmion also has the same topological property \cite{he2022towards,shen2024optical}. With the development of skyrmions in free-space light, this issue becomes increasingly interesting and noteworthy. \par
In this work, the skyrmion field, composed of two Laguerre-Gaussian (LG) modes with orthogonal polarizations, is transmitted through the atmosphere. We numerically demonstrate that the skyrmion number has a certain robustness during propagation in atmospheric turbulence, despite the distortion of the intensity and phase patterns of the field. \par

The article is structured as follows. In Sec. \ref{sec:skyrmioninc beams}, we describe the form and composition of skyrmionic beams. Simultaneously, we express the following formulas, the effective magnetization $\boldsymbol{M}$ corresponding to the Poincar\'{e} vector of light and the skyrmion number. In Sec. \ref{sec:propagation and turbulence model}, we use the split-step method to generate a thick random media to simulate the atmospheric turbulence. In Sec. \ref{sec:numerical results}, we first introduce how turbulence impacts on the intensity and phase patterns of OAM modes and loses the information carried by OAM modes. However, some special structured optical fields, such as skyrmion fields, have topological protection and their skyrmion numbers are rather robust against turbulence, which makes the transmission of information in atmospheric interference possible. In addition, we numerically show that a larger difference between absolute values of azimuthal indices of two spatial modes of the vectorial field results in a superior level of robustness. The conclusions of this paper are given in Sec. \ref{sec:conclusion}.

\section{Skyrmionic beams}
\label{sec:skyrmioninc beams}
\noindent A skyrmionic beam is a kind of vector vortex beam with distinct topological properties and it can be obtained by varying the polarization and field amplitude spatially. We consider a paraxial optical beam and express its unnormalized local state as \cite{gao2020paraxial}
\begin{equation}
    \ket{\widetilde{\Phi}(\boldsymbol{r})} = u_0(\boldsymbol{r})\ket{\varphi} + \exp(i\theta_0)u_1(\boldsymbol{r})\ket{\varphi_\perp},
    \label{state1}
\end{equation}
where $\ket{\varphi}$ and $\ket{\varphi_\perp}$ represent any two orthogonal polarization states, $u_0(\boldsymbol{r})$ and $u_1(\boldsymbol{r})$ are two orthogonal spatial modes, here we only consider the LG modes with respective different azimuthal indices $l_1$ and $l_2$ and the same radial index $p=0$ as the selected spatial modes. $\theta_0$ is the relative phase between two polarization components.\par
In general, in cylindrical coordinates $(r,\phi,z)$, where $r$ is the transverse coordinate component, $\phi$ is the azimuthal component, and $z$ is the propagation distance along the propagation axis, the $LG_{pl}$ mode can be described as \cite{allen1992orbital,allen2016optical,yao2011orbital}
\begin{align}
    u_{pl}(r,\phi,z) = &\frac{C_{pl}}{w(z)}\left [\frac{\sqrt{2}r}{w(z)} \right]^{\mid l \mid}L_p^{\mid l \mid }\left[\frac{2r^2}{w^2(z)}\right]\exp\left[-\frac{r^2}{w^2(z)}\right] \exp(il\phi)
    \notag
    \\& \times \exp\left[\frac{-ikr^2z}{2(z^2+z_R^2)}\right]\exp\left[i(2p+\mid l \mid +1)\arctan\left(\frac{z}{z_R}\right){}\right],
    \label{LGmode}
\end{align}
where $C_{pl} = \sqrt{\frac{2p!}{\pi(p+\mid l \mid)!}}$  is a normalization constant, $L_p^{\mid l \mid}(x)$ is the generalized Laguerre polynomial, $z_R = \frac{\pi w_0^2}{\lambda}$ is the Rayleigh distance, $\lambda$ represents the beam wavelength, $w_0$ is the beam waist of the fundamental mode, and $w(z) = w_0 \sqrt{1+\left(\frac{z}{z_R}\right)^2}$ is the beam radius at the $z$ plane.\par
For a magnetic skyrmion, a normalized local magnetization vector $\boldsymbol{m}$ is the unit vector that represents the direction of the magnetic moments at a specific point in the material and plays a crucial role in describing the magnetic configuration of the skyrmion. Similarly, for an optical skyrmion, it can be well defined by the effective magnetization $\boldsymbol{M}$ which indicates the local direction of the Poincar\'{e} vector. \par
Locally normalizing the state in Eq. (\ref{state1}), we can obtain the form of
\begin{equation}
    \ket{\Phi(\boldsymbol{r})} = \frac{u_0(\boldsymbol{r})\ket{\varphi} + \exp(i\theta_0)u_1(\boldsymbol{r})\ket{\varphi_\perp}}{\sqrt{\mid u_0(\boldsymbol{r}) \mid ^2 + \mid u_1(\boldsymbol{r}) \mid ^2}}
    \label{state2}.
\end{equation}
Therefore, the effective magnetization $\boldsymbol{M}$ corresponding to the local Poincar\'{e} vector of light is
\begin{equation}
    \boldsymbol{M} = \langle \Phi(\boldsymbol{r})|\boldsymbol{\sigma}|\Phi(\boldsymbol{r}) \rangle .\label{M vector}
\end{equation}
Here, $\boldsymbol{\sigma}$ denotes the Pauli operator containing three components $\sigma_x, \sigma_y, \sigma_z$ \cite{gao2020paraxial} and $\sigma_i = \ket{\lambda_i^+}\bra{\lambda_i^+} - \ket{\lambda_i^-}\bra{\lambda_i^-} (i = x,y,z)$, where $\ket{\lambda_i^\pm}$ are the eigenstates corresponding to the eigenvalues $\lambda_i^\pm=\pm 1$ of $\sigma_i$. \par
The indicator of the skyrmion field is the skyrmion number \cite{gao2020paraxial,shen2024optical,liu2022disorder,ornelas2024non,shen2024topologically}, 
\begin{equation}
    N_z = \frac{1}{4\pi}\iint \Sigma_z(x,y) {\rm d}x{\rm d}y
    \label{skyrme number},
\end{equation}
which is employed to quantify the features of the optical field and given by an integral over  a two-dimensional plane. Here the $z$ component of this skyrmion field is
\begin{equation}
    \Sigma_z = \Sigma_{i=z} = \frac{1}{2}\epsilon_{ijk}\epsilon_{pqr}M_p\frac{\partial M_q}{\partial x_j}\frac{\partial M_r}{\partial x_k},
\end{equation}
i.e.,
\begin{align}
     \Sigma_z(x,y) = &\frac{1}{2}\epsilon_{pqr}M_p\frac{\partial M_q}{\partial x}\frac{\partial M_r}{\partial y}
    \notag
    \\&- \frac{1}{2}\epsilon_{pqr}M_p\frac{\partial M_q}{\partial y}\frac{\partial M_r}{\partial x},
\end{align}
where the notations $\epsilon_{ijk}$ and $\epsilon_{pqr}$ are both Levi-civita symbols. According to the forms of Eq. (\ref{state2}) and the vector vortex beam, Eq. (\ref{skyrme number}) can be written as $N_z = \mid l_2 - l_1 \mid$ with $l_1 \neq -l_2$. Thus the skyrmion number is associated with the two orthogonal components of the skyrmion field and theoretically expressed by the difference between two topological charge numbers for ideal LG modes in Eq. (\ref{LGmode}). \par
Here, we should note that $u_0$ and $u_1$ are LG modes, which are not the unique form to construct the skyrmions. In fact, it is only a special form. If both $u_0$ and $u_1$ are in superposition of LG modes, we may still obtain a skyrmion structure with the determinate skyrmion number. When a skyrmionic beam propagates along the \textit{z}-axis in free space, the skyrmion number is unchanged \cite{gao2020paraxial}. Moreover, if the skyrmionic beam passes through a random medium with certain thickness, such as turbulent atmosphere, we need to investigate whether the skyrmion number $N_z$ is stable. In the following, we will use the Kolmogorov theory of turbulence to describe the propagation process from the sender to the receiver and evaluate the robustness of skyrmionic beams after propagating in the turbulent atmosphere.

\section{Propagation and turbulence model}
\label{sec:propagation and turbulence model}
\noindent The atmospheric turbulence phenomenon refers to the inhomogeneous random distribution of the density and refractive index of the atmosphere around the earth due to nonuniform changes of temperature and pressure. Turbulence has a great influence on the propagation of light and distorts the wavefront, so it is vital to understand the specific behaviour of atmospheric turbulence.\par

\subsection{Variation of refractive index}
\noindent In essence, the main factor giving rise to the fluctuations of the refractive index is temperature. The stochastic distribution of refractive index can be mathematically expressed as $n = n_0 + \delta n$, where $n_0$ is the unperturbed term with $n_0 = \langle n \rangle \cong 1$, and the perturbed term $\delta n$ is small, satisfying $\mid\delta n\mid \ll 1 $ and has a zero mean value, i.e., $\langle \delta n \rangle = 0$. Thus, the stochastic paraxial Helmholtz equation for a field $U(x,y,z)$ is as follows \cite{klug2023robust}
\begin{equation}
    \nabla_\perp^2 U + 2ik\frac{\partial U}{\partial z} + 2k^2 \delta n U = 0 
    \label{Helmholtz eqs},
\end{equation}
where $\nabla_\perp^2$ is the transverse Laplacian operator, and $k = 2\pi/\lambda$ is the wave number. The solution of Eq. (\ref{Helmholtz eqs}) is the perturbed complex amplitude and this result can be obtained numerically by employing the split-step method to construct a turbulence model.\par

\subsection{Split-step method with subharmonics}
\noindent For simplicity, we consider linear and isotropic atmospheric turbulence. Turbulence can be appropriately simulated by placing a random phase screen at intervals of $\Delta z = Z/N$ along the propagation path, where $Z$ is the total distance of the turbulence medium and $N$ is the number of random phase screens. This turbulence model is built up by the well-known spilt-step Fourier method which consists of two periodic alternating propagation processes named diffraction and refraction, respectively. The free-space diffraction result is in accordance with the Huygens–Fresnel integral in vacuum and the refraction is described by a random phase factor $\Theta_j$ resulting from the inhomogeneous refractive index $\delta n$, where $j$ labels the $j$th screen \cite{schmidt2010numerical}.\par
\begin{figure}[t]
    \centering
    \subfigure[]{
    \begin{minipage}[t]{0.7\linewidth}
        \centering
        \includegraphics[width=3.8in]{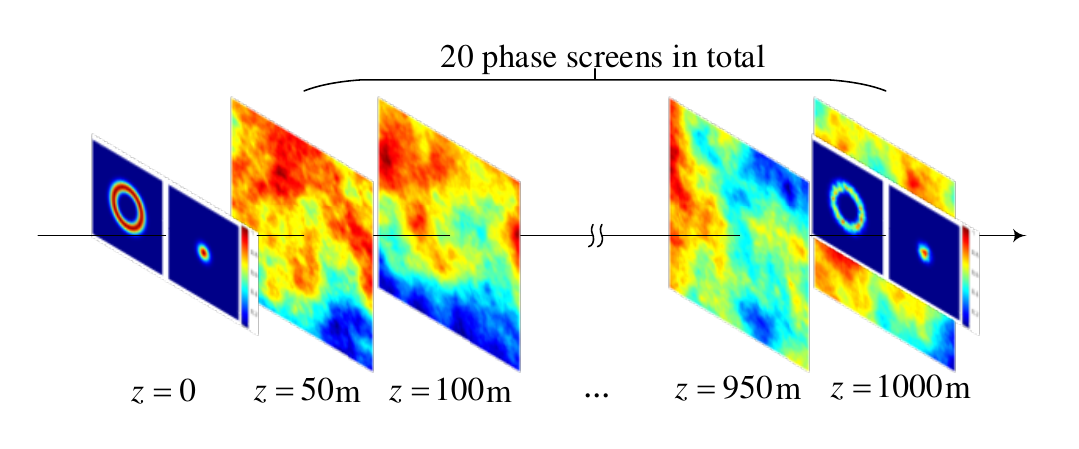}
        \label{fig:turbulence_model}
    \end{minipage}
    }\subfigure[]{
    \begin{minipage}[t]{0.3\linewidth}
        \centering
        \includegraphics[width=1.5in]{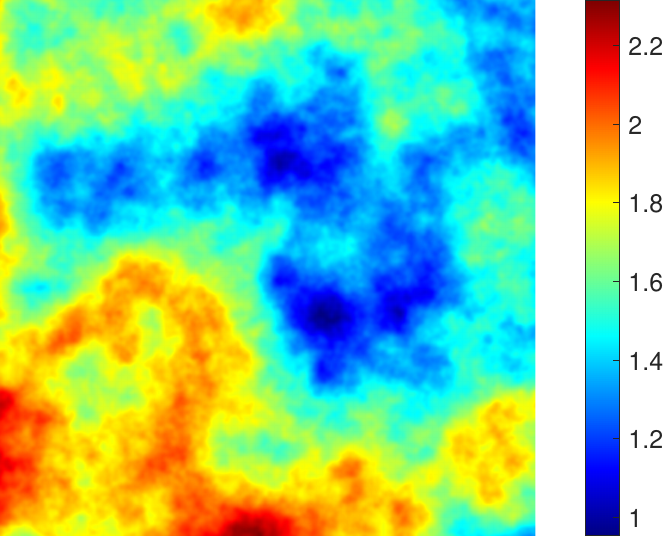}
        \label{fig:phz_ft_sh1}
    \end{minipage}
    }
    \caption{(a) Schematic of the turbulence model with a split-step method. (b) A random phase screen example of turbulence created by the FFT method with the addition of subharmonics.}
\end{figure}

As shown in Fig. \ref{fig:turbulence_model}, the total length of the atmosphere is $Z = 1$km, and we split it into $N = 20$ thin random screens separated by $\Delta z = 50$m. The generation of a single random screen at each plane corresponds to the modified spectrum of refractive index \cite{andrews2005laser}
\begin{align}
    \Psi_n(\kappa) =  &0.033C_n^2(\kappa^2 + \kappa_0^2)^{-11/6}\exp(-\kappa^2/\kappa_\mathcal{l}^2)
    \notag
    \\& \times \left[1 + 1.802(\kappa/\kappa_\mathcal{l}) - 0.254(\kappa/\kappa_\mathcal{l})^{7/6}\right],
\end{align}
where $C_n^2$ is the structure constant of the refractive index associated with the turbulence strength, $\kappa$ is the spatial frequency in units of rad/m, $\kappa_\mathcal{l} = 3.3/\mathcal{l}_0$, $\kappa_0 = 2\pi/L_0$, $\mathcal{l}_0$ is the turbulence inner scale, and $L_0$ is the turbulence outer scale. The screen interval, denoted by $\Delta z$, must be larger than the maximum atmospheric heterogeneity $L_0$, to uphold the Markov assumption. However, it should also be sufficiently small to ensure the rationality of the model's propagation from the $z_j$ to $z_{j+1}$ plane \cite{belmonte2000feasibility}. Under the condition of the plane wave approximation, another characteristic parameter of turbulence strength is the Fried parameter $r_0$ \cite{gu2020phenomenology,doster2016laguerre}, which is defined as follows
\begin{equation}
    r_0 = (0.423k^2zC_n^2)^{-3/5},
\end{equation}
where $z$ is the distance of turbulence covered by the structure constant $C_n^2$.
Further, the power spectrum function of the phase can be expressed as \cite{belmonte2000feasibility}
\begin{align}
    \Psi_\theta(\kappa) =  &2\pi k^2z\Psi_n(\kappa)
    \notag
    \\=  &0.490r_0^{-5/3}(\kappa^2 + \kappa_0^2)^{-11/6}\exp(-\kappa^2/\kappa_\mathcal{l}^2)
    \notag
    \\ &\times \left[1 + 1.802(\kappa/\kappa_\mathcal{l}) - 0.254(\kappa/\kappa_\mathcal{l})^{7/6}\right].
\end{align}
A single random phase screen can be created by using the fast Fourier transform (FFT) method \cite{McGlamery:67} of the Gaussian complex random variables with a zero mean value and unit variance modulated by the power spectrum function of the phase $\Psi_\theta(\kappa)$. \par
However, when there are large scale phase fluctuations, the lower frequency part exists substantial effects and non-negligible errors for screens because of the aperture, i.e., these frequencies in the spectral region $(-\kappa_{x,y}/2, \kappa_{x,y}/2)$, where $\kappa_{x,y} = 2\pi/D_{x,y}$, $\kappa_x^2 + \kappa_y^2 = \kappa^2$, and $D_{x,y}$ is the size of the screen. Therefore, the addition of subharmonics introduced by Lane \textit{et al.} is essential to align with the more sophisticated turbulent flow theory, which more accurately reflects the complexities inherent in real-world turbulence scenarios \cite{lane1992simulation,frehlich2000simulation}. 
A random screen example with $C_n^2 = 1\times 10^{-15}\mathrm{m}^{-2/3}$ and $z = 50$m at wavelength $\lambda = 780$nm is shown in Fig. \ref{fig:phz_ft_sh1}. It should be noted that the selection of the wavelength does not affect this model and simulation, although other wavelengths are often considered in free space communication.

\section{Numerical results: Non-trivial topological property of skyrmions}
\label{sec:numerical results}
\noindent We choose a numerical grid size of $1024\times 1024$, and the screen is a square region with $D = D_{x,y} = 0.5$m, the wavelength is $\lambda = 780$nm, and the beam waist is $w_0 = 5$cm at $z_0=0$ plane, the turbulence inner and outer scales are $\mathcal{l}_0 = 5$mm and $L_0 = 20$m respectively. In this paper, we select four distinct structure constants, denoted as $C_n^2 = \{ 1\times 10^{-16}\mathrm{m}^{-2/3}, 1\times 10^{-15}\mathrm{m}^{-2/3}, 1\times 10^{-14}\mathrm{m}^{-2/3}, 3\times 10^{-14}\mathrm{m}^{-2/3}\}$, to represent the varying degrees of turbulence across the entire journey. And these constants correspond to turbulence strengths that range from weak to strong \cite{doster2016laguerre}. Furthermore, we perform 1000 realizations for the ensemble average on account of the existence of stochastic process and this averaging operation is performed on the final results, denoting multiple trials.

\subsection{Perturbations of LG modes}
\begin{figure*}[t]
    \centering
    \subfigure{
    \begin{minipage}[t]{0.49\linewidth}
        \centering
        \includegraphics[width=2.9in]{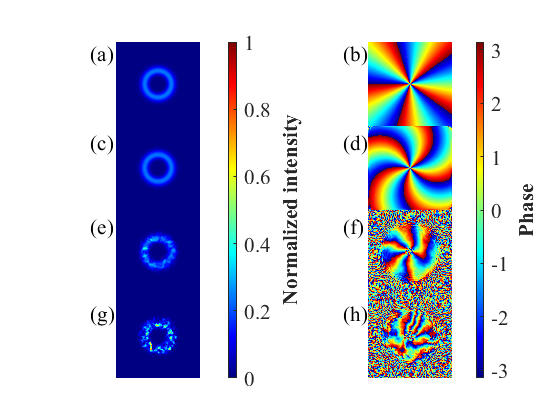}
    \end{minipage}
    }\hspace{-5mm}
    \subfigure{
    \begin{minipage}[t]{0.49\linewidth}
        \centering
        \includegraphics[width=2in]{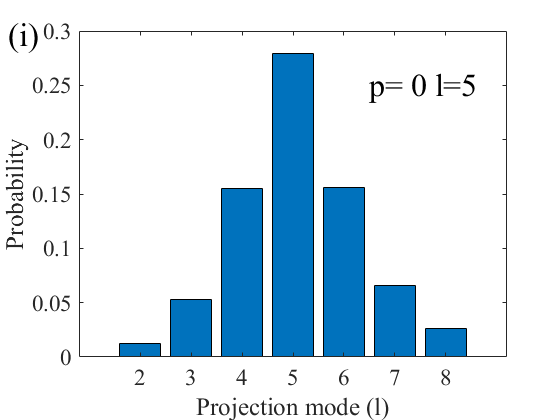}
        \label{P=0&L=5_ProjectionProb_Cn2_-15}
    \end{minipage}
    }
    \caption{Intensity and phase distributions for the LG$_{05}$ mode when propagating without and with turbulence. (a), (b): The intensity and phase patterns of the initial input beam. (c), (d): The intensity and phase patterns after passing through the free space with no turbulence. (e), (f) and (g), (h): The intensity and phase patterns acquired by the receiver at $z = 1$km plane after propagating in different turbulence ($C_n^2 = 1\times 10^{-15} \mathrm{m}^{-2/3}$ and $C_n^2 = 1\times 10^{-14} \mathrm{m}^{-2/3}$, respectively). Colorbar: the normalized intensity and phase ranging in $[-\pi, \pi]$. (i): The projection probability distribution of the perturbed OAM beam corresponding to $C_n^2 = 1\times 10^{-15} \mathrm{m}^{-2/3}$. The initial topological charge of the beam is $l=5$ with radial quantum number $p=0$, and the projected modes have the same radial indices $p=0$ and different azimuthal indices $l$ (here $l$ is not limited these seven numbers and we have just shown some of them).}
    \label{fig:LG05patterns}
\end{figure*}

\noindent In our study, we consider the LG spatial modes as the degree of freedom of space for the purpose of coupling with the polarization to construct a desired skyrmion field. Hence, we can first observe several types of patterns of LG modes in turbulence. For instance, we take the LG$_{05}$ ($p=0, l=5$) mode that is subjected to the turbulence characterized by $C_n^2 = 1\times 10^{-15}\mathrm{m}^{-2/3}$ and $1\times 10^{-14}\mathrm{m}^{-2/3}$. The numerical results of intensity and phase are shown in Fig. \ref{fig:LG05patterns}. \par

The two pictures in the top row represent the intensity and phase of the initial LG$_{05}$ mode, the second row represents the free-space output patterns, and the last two rows represent the patterns after propagating in turbulence. They indicate that atmospheric turbulence has an obvious impact on propagating LG modes that carry OAM while those patterns in which the beam propagates through a channel with no turbulence maintain the primary spiral forms, with the exception of an additional phase attributed to the Gouy phase shift \cite{feng2001physical,hamazaki2006direct} and wavefront curvature. Particularly, it is worth noting that the phase profiles are profoundly sensitive to the presence of turbulence. As the turbulence escalates, it inevitably leads to the disruption of the corresponding phase vortex. Additionally, the topological charge of an OAM beam propagating through atmospheric turbulence can evolve into a superposition of diverse optical modes characterized by distinct topological charges, as illustrated in Fig. \ref{fig:LG05patterns}(i). This demonstrates the instability of OAM modes in turbulent environments and the potential loss of information that ensues. Nevertheless, skyrmions exhibit a different nature.\par

\subsection{Perturbations of skyrmion field}
\noindent In Section 2, we have mentioned that a skyrmionic beam is composed of two orthogonal polarized LG spatial beams. Since a skyrmion field is launched into atmospheric turbulence, each component of the vector field will be perturbed and both the intensity and phase patterns will be disturbed to some extent. However, our main concern is the variation of  the skyrmion number.\par
To obtain the effective magnetization $\boldsymbol{M}$ with three components corresponding to the expected values of the Pauli operators in the normalized local state, the post-selection results for six polarization bases $\{\ket{D}, \ket{A}, \ket{L}, \ket{R},$ $\ket{H}, \ket{V}\}$ which are the eigenstates of the Pauli operators are needed, where $\ket{D} = \frac{1}{\sqrt{2}}(\ket{H} + \ket{V}), \ket{A} = \frac{1}{\sqrt{2}}(\ket{H} - \ket{V}), \ket{L} = \frac{1}{\sqrt{2}}(\ket{H} + i\ket{V}), \ket{R} = \frac{1}{\sqrt{2}}(\ket{H} - i\ket{V}$ ($\ket{H}$ and $\ket{V}$ represent the horizontal polarization and vertical polarization, respectively). Because we express $\boldsymbol{M}$ in Eq. (\ref{M vector}) in a locally normalized state by $M_i(\boldsymbol{r}) = \langle \sigma_i(\boldsymbol{r}) \rangle = \frac{I_i^+(\boldsymbol{r})-I_i^-(\boldsymbol{r})}{I_i^+(\boldsymbol{r})+I_i^-(\boldsymbol{r})}$, where $i = x,y,z$ and $I_i^\pm(\boldsymbol{r}) = |\langle\lambda_i^\pm|\Phi(\boldsymbol{r})\rangle|^2$ are the projection intensities of the state $\Phi(\boldsymbol{r})$ on the eigenstates $\ket{\lambda_i^\pm}$ related to the eigenvalues $\lambda_i^\pm = \pm1$ of the Pauli operator $\sigma_i$ ($\sigma_x = \ket{D}\bra{D}-\ket{A}\bra{A} = \ket{H}\bra{V}+\ket{V}\bra{H}$, $\sigma_y = \ket{L}\bra{L}-\ket{R}\bra{R} = -i\ket{H}\bra{V}+i\ket{V}\bra{H}$ and $\sigma_z = \ket{H}\bra{H}-\ket{V}\bra{V}$). Hence, by the above $\boldsymbol{M}$ vectors in each transverse position $\boldsymbol{r}$, the corresponding topological structure of the skyrmion is established, along with the relevant skyrmion number. \par 

\begin{figure*}[t]
    \centering
    \subfigure[$C_n^2 = 1\times 10^{-16}\mathrm{m}^{-2/3}$]{
    \begin{minipage}[t]{0.49\linewidth}
        \centering
        \includegraphics[width=1.8in]{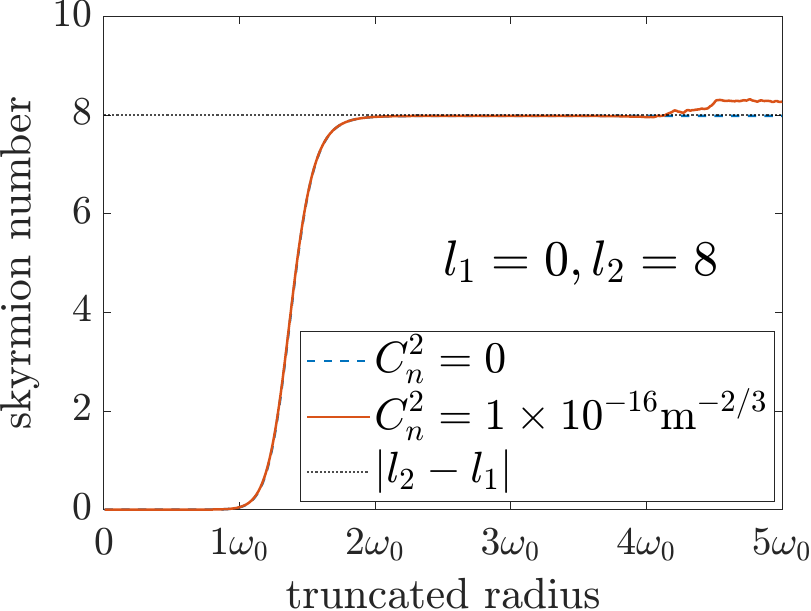}
        \label{L0_8_001}
    \end{minipage}
    }\subfigure[$C_n^2 = 1\times 10^{-15}\mathrm{m}^{-2/3}$]{
    \begin{minipage}[t]{0.49\linewidth}
        \centering
        \includegraphics[width=1.8in]{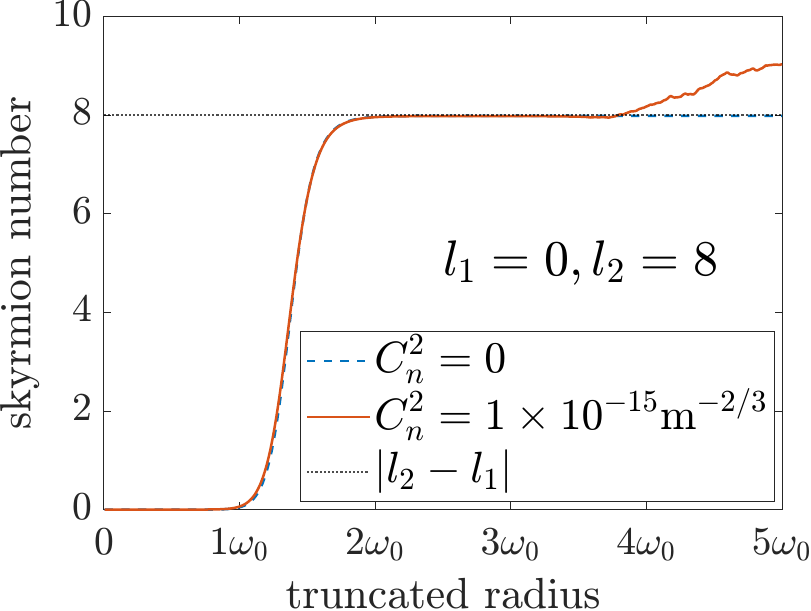}
        \label{L0_8_002}
    \end{minipage}
    }
    
    \subfigure[$C_n^2 = 1\times 10^{-14}\mathrm{m}^{-2/3}$]{
    \begin{minipage}[t]{0.49\linewidth}
        \centering
        \includegraphics[width=1.8in]{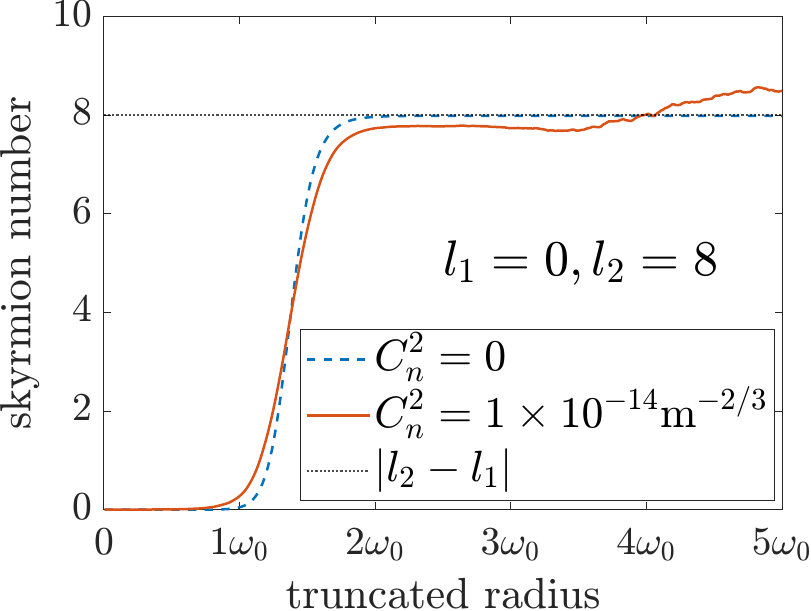}
        \label{L0_8_003}
    \end{minipage}
    }\subfigure[$C_n^2 = 3\times 10^{-14}\mathrm{m}^{-2/3}$]{
    \begin{minipage}[t]{0.49\linewidth}
        \centering
        \includegraphics[width=1.8in]{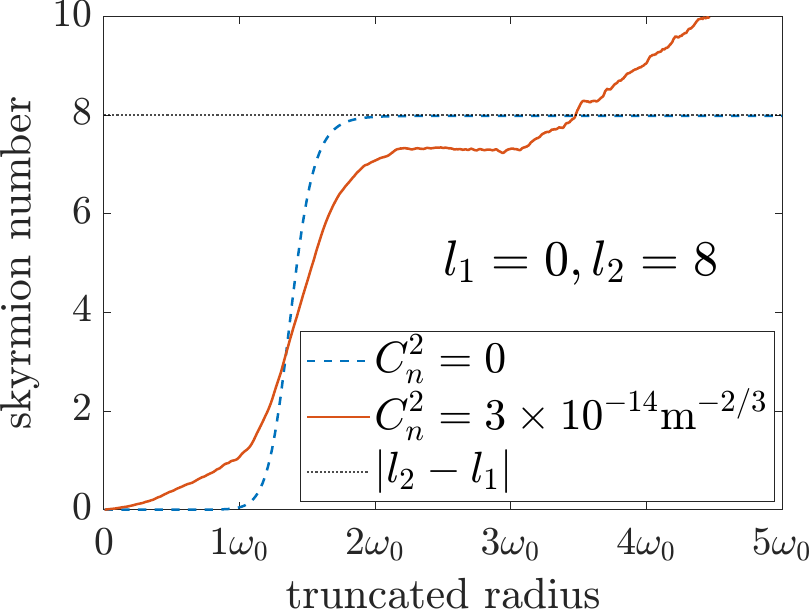}
        \label{L0_8_004}
    \end{minipage}
    }
    \caption{The curves of skyrmion numbers with the truncated radius. The black dotted line: the value $|l_2-l_1|$. The blue solid line: the skyrmion numbers in free space. The orange solid line: the skyrmion numbers in turbulence. (a)-(d): two OAM modes $l_1 = 0, l_2 = 8$, the turbulence refractive index constants $C_n^2 = \{1\times 10^{-16}\mathrm{m}^{-2/3}, 1\times 10^{-15}\mathrm{m}^{-2/3}, 1\times 10^{-14}\mathrm{m}^{-2/3}, 3\times 10^{-14}\mathrm{m}^{-2/3}\}$.}
    \label{fig:truncation_6}
\end{figure*}

It can be seen in Eq. (\ref{skyrme number}) that the calculation of skyrmion numbers requires integration over the whole plane. However, we need to implement a proper truncation of the integral region to mitigate the significant fluctuation of skyrmion numbers caused by regions with very low intensities. The intensity of the outer region of the spot decreases exponentially with the increase of the distance from the center, making it negligible. And this low intensity can have a detrimental effect on the gradient of the effective magnetization, leading to significant disturbances in the skyrmion numbers. Thus, the truncation of the integral area is necessary. We have illustrated in Fig. \ref{fig:truncation_6} the curves depicting the variation of skyrmion numbers with respect to the truncation radius in several scenarios. The truncation radius is measured in units of the fundamental mode waist radius $w_0$ (the beam waist $w_0 = 5$cm in this paper), and it varies from 0 to 500\% of the waist radius so that the maximum truncated circle can be approximately tangent to the edge of the viewing screen that we selected. Meanwhile the relevant skyrmion numbers over varying circular areas are calculated. We can see that the final skyrmion number is close to the ideal theoretical value $|l_2-l_1|$ and then remains nearly constant as the truncated circle contains a full light spot. Indeed, the optimal selection of the truncated radius involves choosing an appropriate and clearly defined value that encompasses the entire spot and ensures stable data trends for the desired skyrmion number (the more details are in \ref{app:truncation area}). 

\begin{figure}[htbp!]
    \centering
    \includegraphics[width = 5.2in]{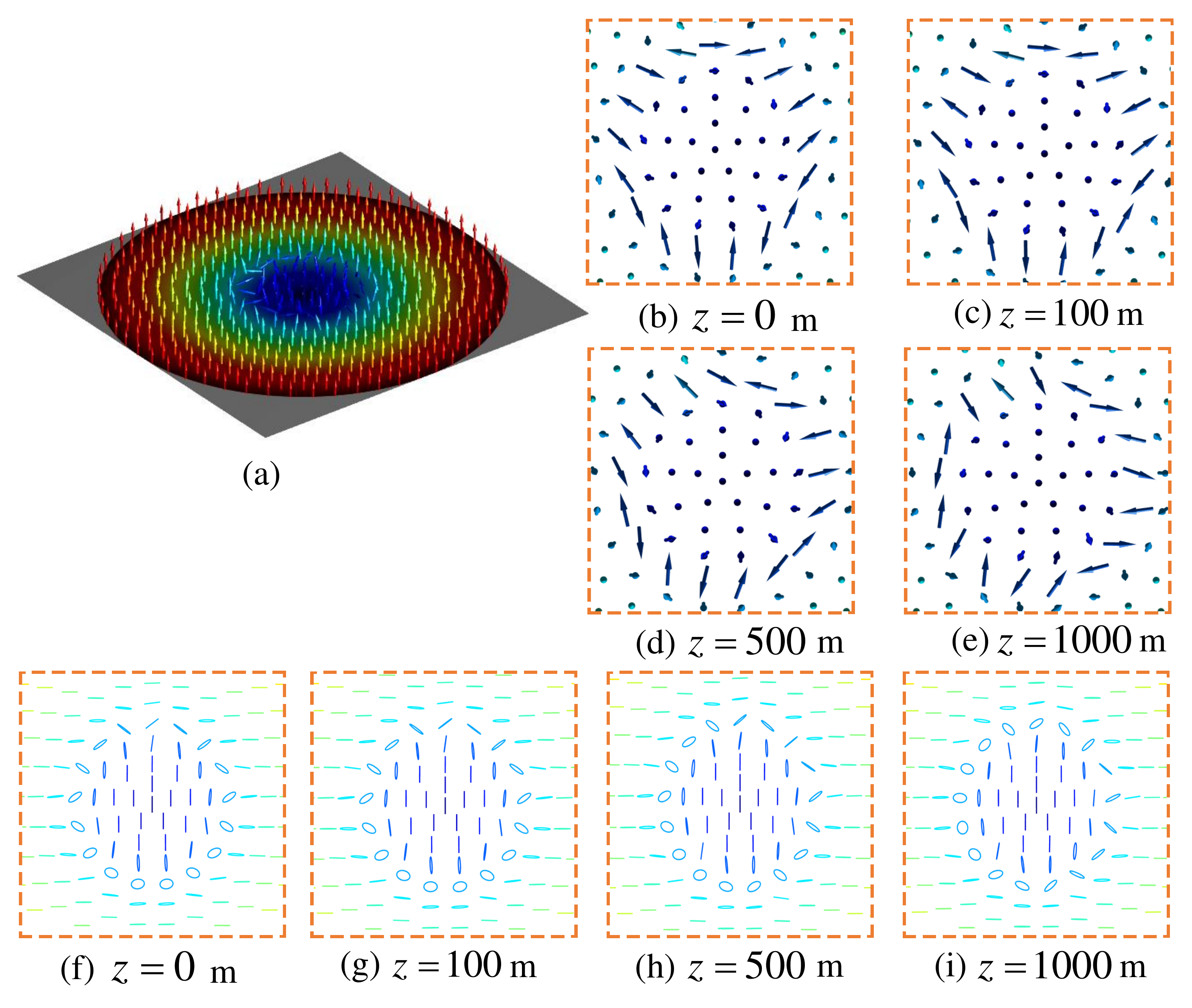}
    \caption{The demonstrations of the skyrmion spin textures (or effective magnetizations) and polarization structures during the propagation. We choose an initial skyrmion number of $8$ constructed by two spatial components ($l_1 = 0, l_2 = 8$) with two orthogonal polarizations as an example. (a): The initial skyrmion spin configuration and its effective magnetization vary with spatial positions. (b)-(e): The strength of this atmospheric turbulence is $C_n^2 = 10^{-15}\mathrm{m}^{-2/3}$. We show the effective magnetizations at four planes ($z=0, 100\mathrm{m}, 500\mathrm{m}, 1000\mathrm{m}$) in the propagation path. To clearly observe the changes of the directions of spin vectors along with the turbulence, (b), (c), (d) and (e) respectively represent the central parts of what we intercepted and enlarged. We only need to pay attention to the number of times of the period of the vector changes in a certain circle to obtain the corresponding skyrmion number (go around and change the arrow $N=8$ times). (f)-(g): The spatially varying polarization ellipses corresponding to (b)-(e).}
    \label{fig:skyrmion_spin}
\end{figure}

In Fig. \ref{L0_8_001}-\ref{L0_8_004}, the two orthogonal modes of the simulated skyrmion fields are $l_1 = 0, l_2 = 8$ but there are different strengths of turbulence, and the numerically calculated skyrmion numbers after the averaging of the ensemble are $7.9908 \pm 0.0001, 7.9860 \pm 0.0005, 7.7729 \pm 0.5169, 7.2948\pm 3.0066$, respectively (The intensity distributions of the six polarization states corresponding to this vector light field are shown in \ref{six proj bases with 0&8}). We also simulate the case that the initial light field has $l_1=1$ and $l_2=9$ (the more details are in \ref{app:skyrmion field with l1=1&l2=9}). It follows that the skyrmion number holds up well when the structure constant $C_n^2$ of turbulence is less than or equal to $1\times 10^{-14}\mathrm{m}^{-2/3}$. With the turbulence gets stronger, the skyrmion number is slightly affected but still recognizable, as shown in Fig. \ref{L0_8_004}. Subsequently, we also illustrate the topological structure of skyrmions at some typical positions along the propagation path, enhancing intuitive understanding of how turbulence influences skyrmion numbers and textures. As seen in Fig. \ref{fig:skyrmion_spin}, taking moderate turbulence and an initial skyrmion number of $8$ as an example, we provide related skyrmion spin textures (or effective magnetizations) and spatially varying polarization structures at four $z$ planes ($z=0, 100\mathrm{m}, 500\mathrm{m}, 1000\mathrm{m}$). To clearly observe the changes in spin vectors and classify the skyrmion structure to obtain the corresponding skyrmion number, we present enlarged images of the central parts in the $x-y$ transverse plane at four sequential $z$ planes. The skyrmion number is determined by how many times the spin vector changes as the azimuthal angle varies from $0$ to $2\pi$ \cite{nagaosa2013topological,shen2022generation,shen2021supertoroidal}. Those images in Fig. \ref{fig:skyrmion_spin}(b)-\ref{fig:skyrmion_spin}(e) demonstrate that an arrow can change $N=8$ times when making a circle, while $N$ remains constant during the propagation of turbulence. Correspondingly, we can also see that the major-axis of the polarization ellipse does half the rotation of the spin vector \cite{gao2020paraxial}. It means that the topological structure of skyrmions is not damaged and the skyrmion number stays unchanged in this case. The skyrmion numbers in other cases can be similarly analyzed and attain similar conclusions.

\subsection{The topological quantity: skyrmion number}
\begin{figure*}[!htbp]
    \centering
    \subfigure[$l_1=0, l_2>0$]{
    \begin{minipage}[t]{0.5\linewidth}
        \centering
        \includegraphics[width=2.2in]{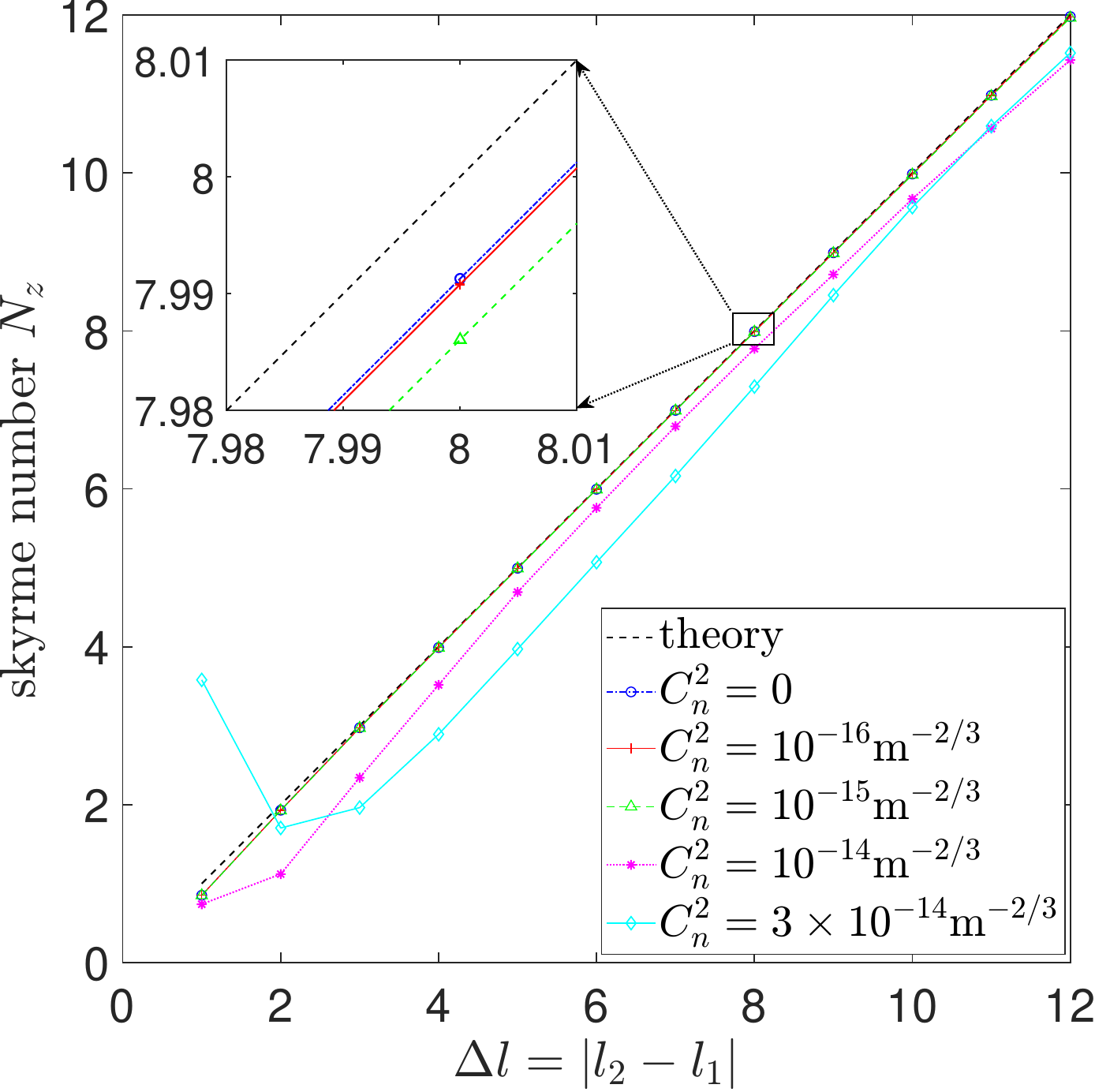}
        \label{fig:L1_0_L2_+_skyrmion_number}
    \end{minipage}
    }\subfigure[$l_1=1, l_2>0$]{
    \begin{minipage}[t]{0.5\linewidth}
        \centering
        \includegraphics[width=2.2in]{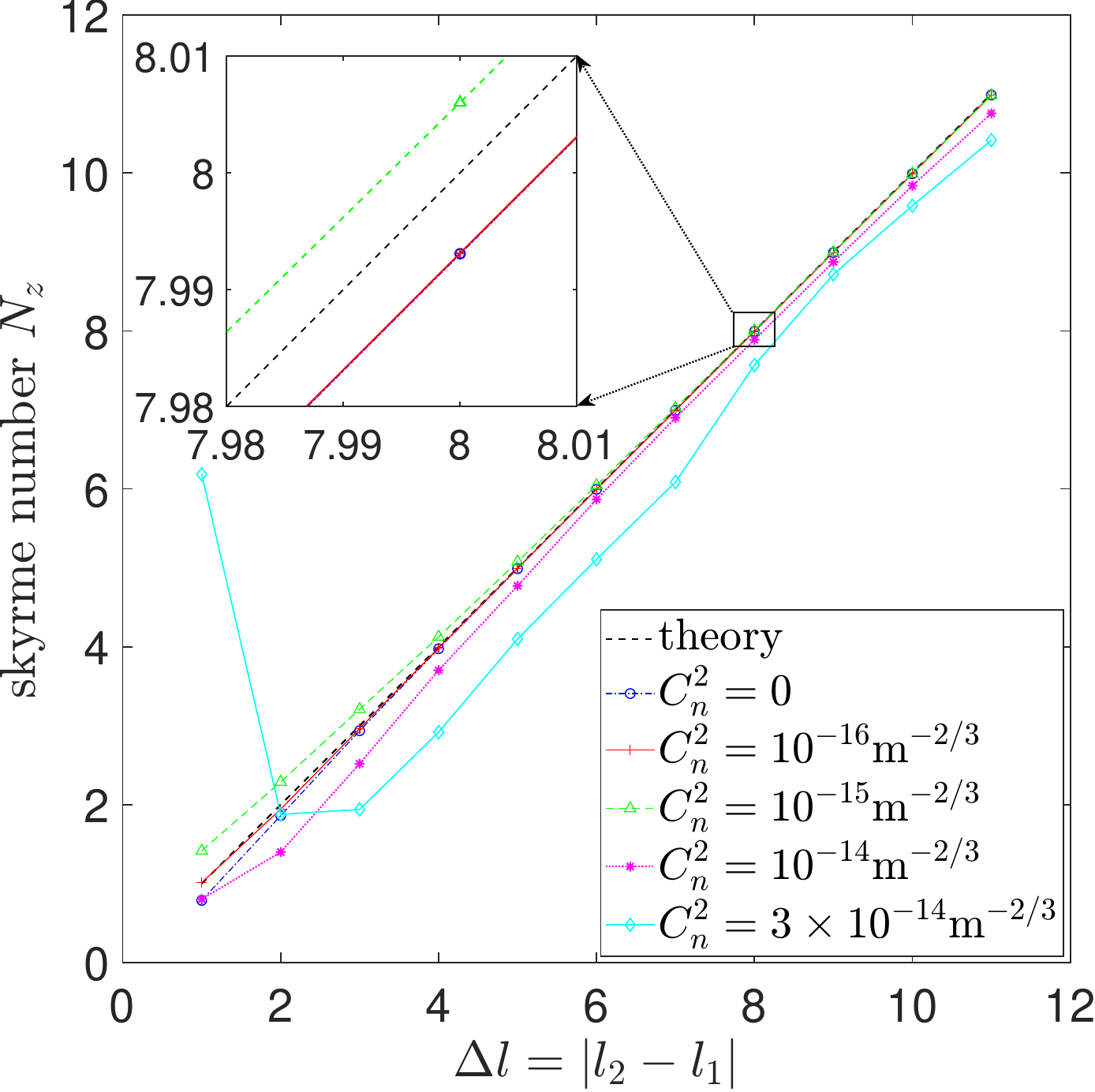}
        \label{fig:L1_1_L2_+_skyrmion_number}
    \end{minipage}
    }
    
    \subfigure[$l_1=-1, l_2>0$]{
    \begin{minipage}[t]{0.5\linewidth}
        \centering
        \includegraphics[width=2.2in]{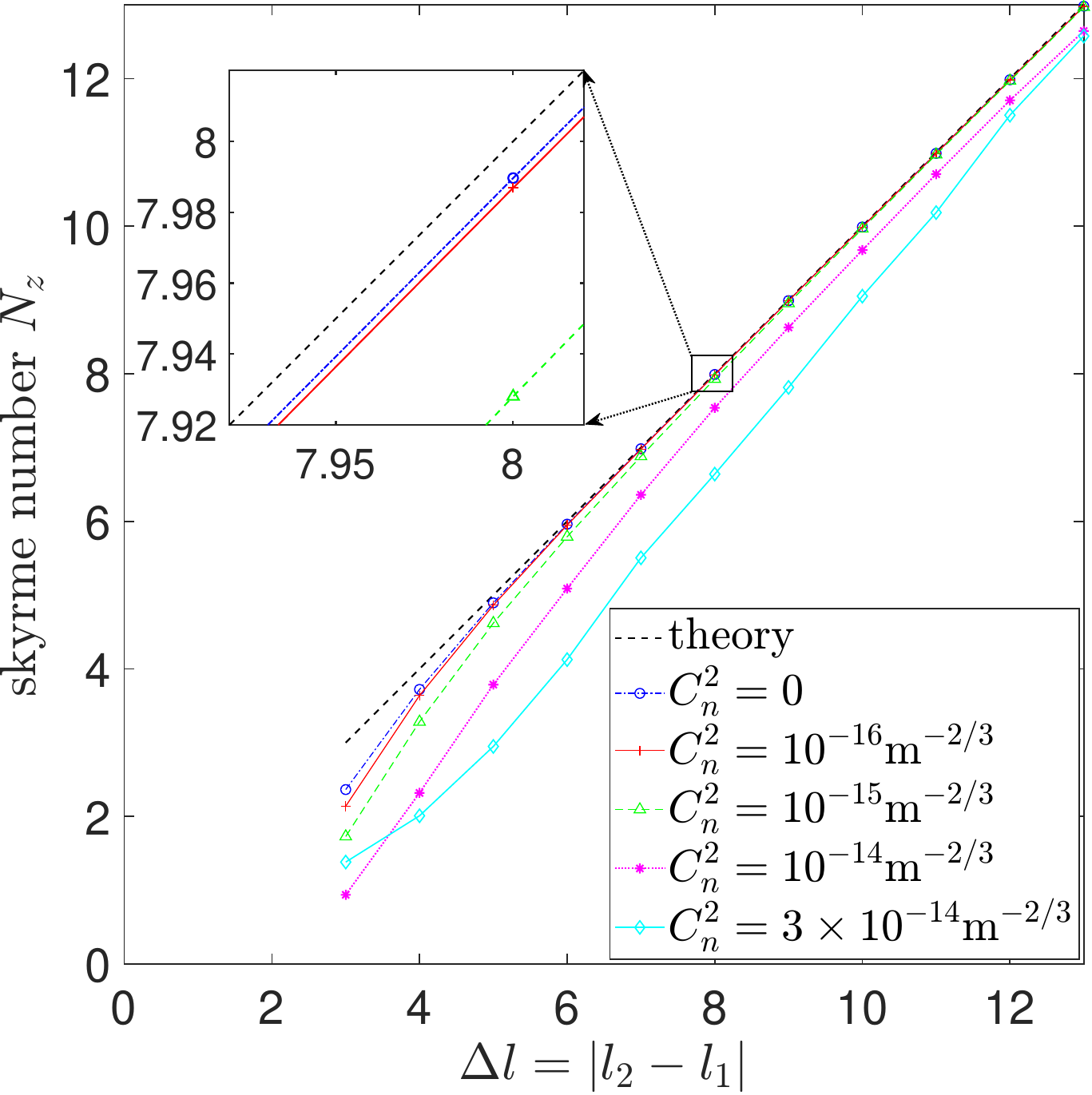}
        \label{fig:L1_-1_L2_+_skyrmion_number}
    \end{minipage}
    }
    \caption{The variation of those skyrmion numbers of multiple skyrmion fields subjected to atmospheric turbulence with different strengths. The five kinds of refractive index structure constants, $C_n^2 = \{0, 10^{-16}\mathrm{m}^{-2/3}, 10^{-15}\mathrm{m}^{-2/3}, 10^{-14}\mathrm{m}^{-2/3}, 3\times10^{-14}\mathrm{m}^{-2/3}\}$, show that the propagation channels of the skyrmionic beam separately are free space (no turbulence) and weak to strong turbulence. Considering the different signs of the azimuthal index $l$ of the OAM mode, we numerically simulate three cases. (a) $l_1=0, l_2>0$. The skyrmionic beam consists of two orthogonal modes with $l_1=0$ and $l_2=\{1,2,3,4,5,6,7,8,9,10,11,12\}$ corresponding to $\Delta l=|l_2-l_1|= \{1,2,3,4,5,6,7,8,9,10,11,12\}$. (b) $l_1=1, l_2>0$. The skyrmionic beam consists of two orthogonal modes with $l_1=1$ and $l_2=\{2,3,4,5,6,7,8,9,10,11,12\}$ corresponding to $\Delta l=|l_2-l_1|= \{1,2,3,4,5,6,7,8,9,10,11\}$. (c) $l_1=-1, l_2>0$. The skyrmionic beam consists of two orthogonal modes with $l_1=-1$ and $l_2=\{2,3,4,5,6,7,8,9,10,11,12\}$ corresponding to $\Delta l=|l_2-l_1|= \{3,4,5,6,7,8,9,10,11,12,13\}$. The insets magnify the data points of $\Delta l=8$.}
    \label{skyrmion_number_curve}
\end{figure*}

\noindent The skyrmion field has the non-trivial topological property that can be described by the robustness of the skyrmion number. To verify this, we numerically research the variations of the skyrmion number in atmospheric turbulence with different strengths. As shown in Fig. \ref{skyrmion_number_curve}, the skyrmion number almost maintains its robustness after passing through the atmospheric turbulence with structure constants of $C_n^2 = \{10^{-16}\mathrm{m}^{-2/3},10^{-15}\mathrm{m}^{-2/3},10^{-14}\mathrm{m}^{-2/3}\}$. When $C_n^2$ reaches the order of $10^{-14}\mathrm{m}^{-2/3}$, this is already a not very small turbulence strength. If we continue to increase the strength to a larger degree, the skyrmion numbers can inevitably be affected due to the stronger turbulence.\par
In Fig. \ref{fig:L1_0_L2_+_skyrmion_number}, \ref{fig:L1_1_L2_+_skyrmion_number}, \ref{fig:L1_-1_L2_+_skyrmion_number}, we choose three different combinations of symbols for $l_1$ and $l_2$ to illustrate the robustness of skyrmion numbers, not only in free space but also in atmospheric turbulence. The black dashed lines represent the theoretical values $N_z = \Delta l = |l_2 - l_1|$, and we should note that on the whole, the skyrmion numbers corresponding to different refractive index structure constants of the atmosphere are close to the theoretical curves when $C_n^2$ is less than or equal to $1\times10^{-14}\mathrm{m}^{-2/3}$. The insets magnify the data points of $\Delta l = 8$. With stronger turbulence, i.e., a higher structure constant $C_n^2$, the propagation of a skyrmionic beam will suffer a greater impact and there will be a larger deviation from the theoretical value of the skyrmion number. When $C_n^2=3\times10^{-14}\mathrm{m}^{-2/3}$, as depicted by the cyan curves with diamond marks in Fig. \ref{skyrmion_number_curve} shows, skyrmion numbers exhibit significant deviations. It is noteworthy that the smaller the difference between absolute values of $l_1$ and $l_2$, the larger the derivation. For instance, the case of $l_1=0$ and $l_2=1$ (or $l_1=1$ and $l_2=2$ or $l_1=-1$ and $l_2=2$) with a difference of absolute values of $1$, is not as stable as in the case of $l_1=0$ and $l_2=11$ (or $l_1=1$ and $l_2=12$ or $l_1=-1$ and $l_2=12$) with a difference of $||l_1|-|l_2||=11$ when the turbulence strength associated with the structure constant $C_n^2$ of refractive index of atmosphere becomes stronger from $10^{-16}\mathrm{m}^{-2/3}$ to $3\times10^{-14}\mathrm{m}^{-2/3}$. The phenomenon may be interpreted by the increased overlap between two beam spots. As $|l_1|$ gets close to $|l_2|$, the overlap is larger and the deleterious effects caused by the turbulence show up more in the skyrmion number related to two components.\par

\renewcommand{\dblfloatpagefraction}{.9}

\begin{figure*}[!htbp]
    \centering
    \subfigure[]{
    \begin{minipage}[t]{0.49\linewidth}
        \centering
        \includegraphics[width=2.2in]{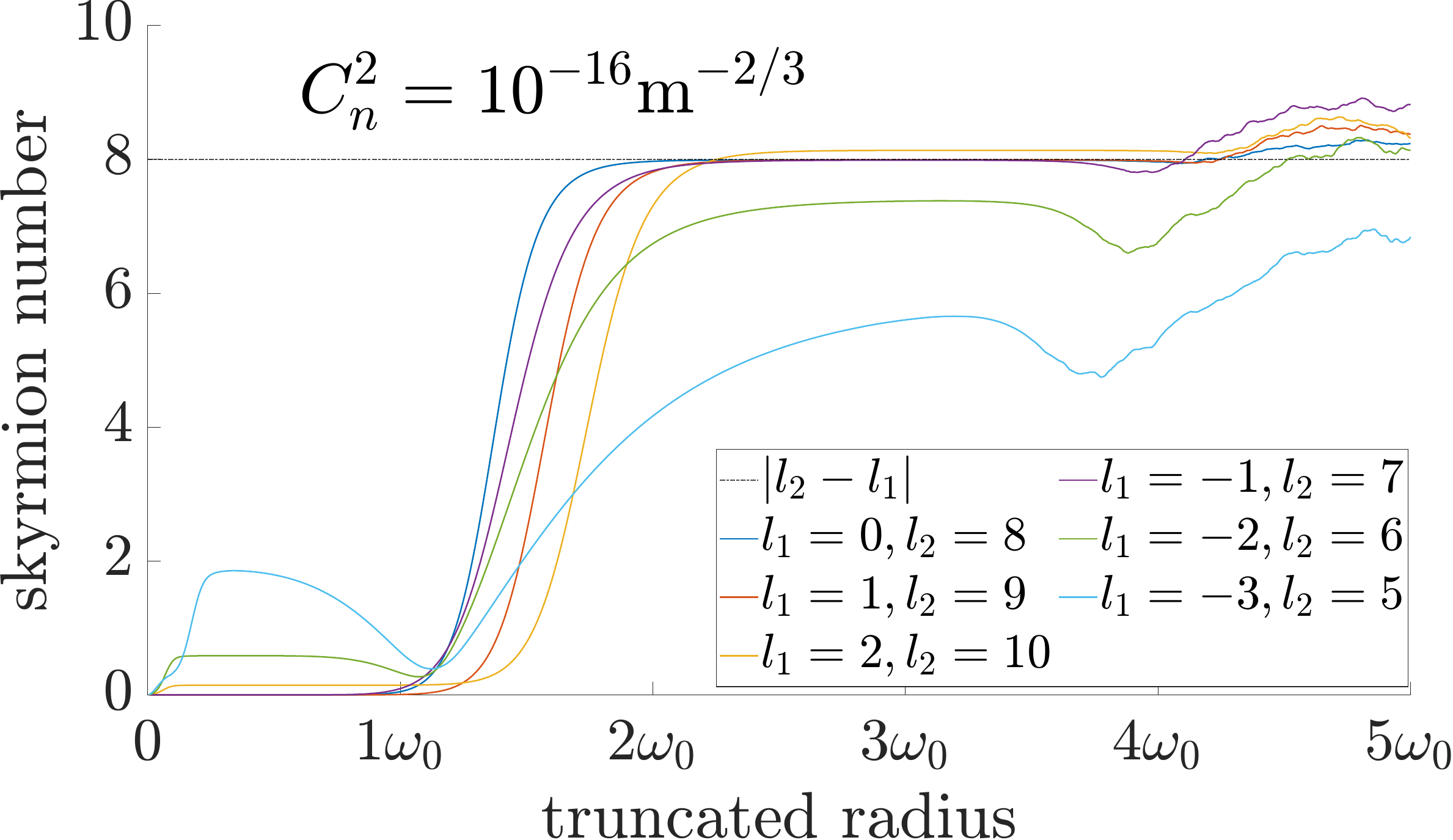}
        \label{fig:Nz_8_Cn2_-16}
    \end{minipage}
    }\subfigure[]{
    \begin{minipage}[t]{0.49\linewidth}
        \centering
        \includegraphics[width=2.2in]{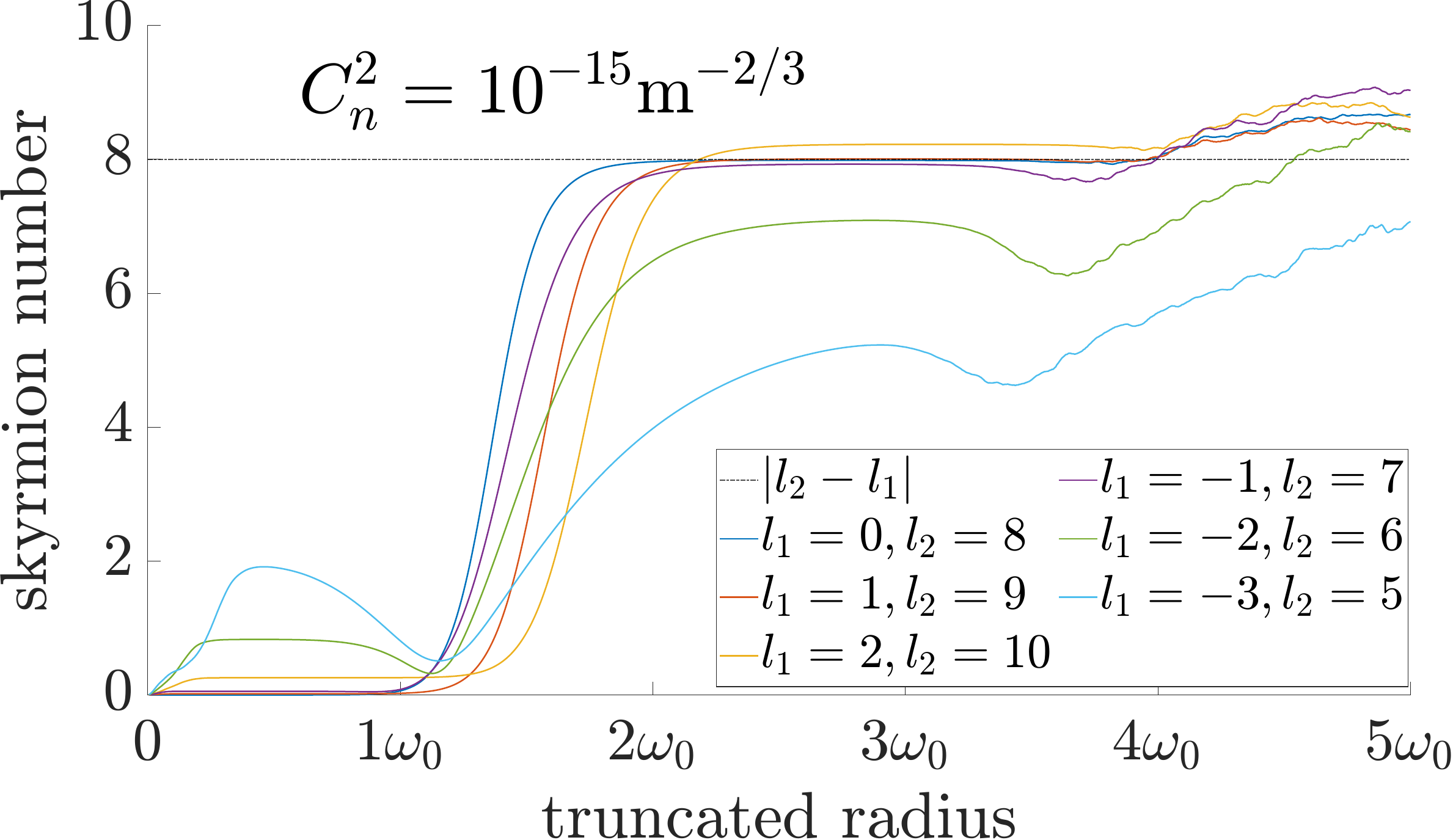}
        \label{fig:Nz_8_Cn2_-15}
    \end{minipage}
    }
    
    \subfigure[]{
    \begin{minipage}[t]{0.49\linewidth}
        \centering
        \includegraphics[width=2.2in]{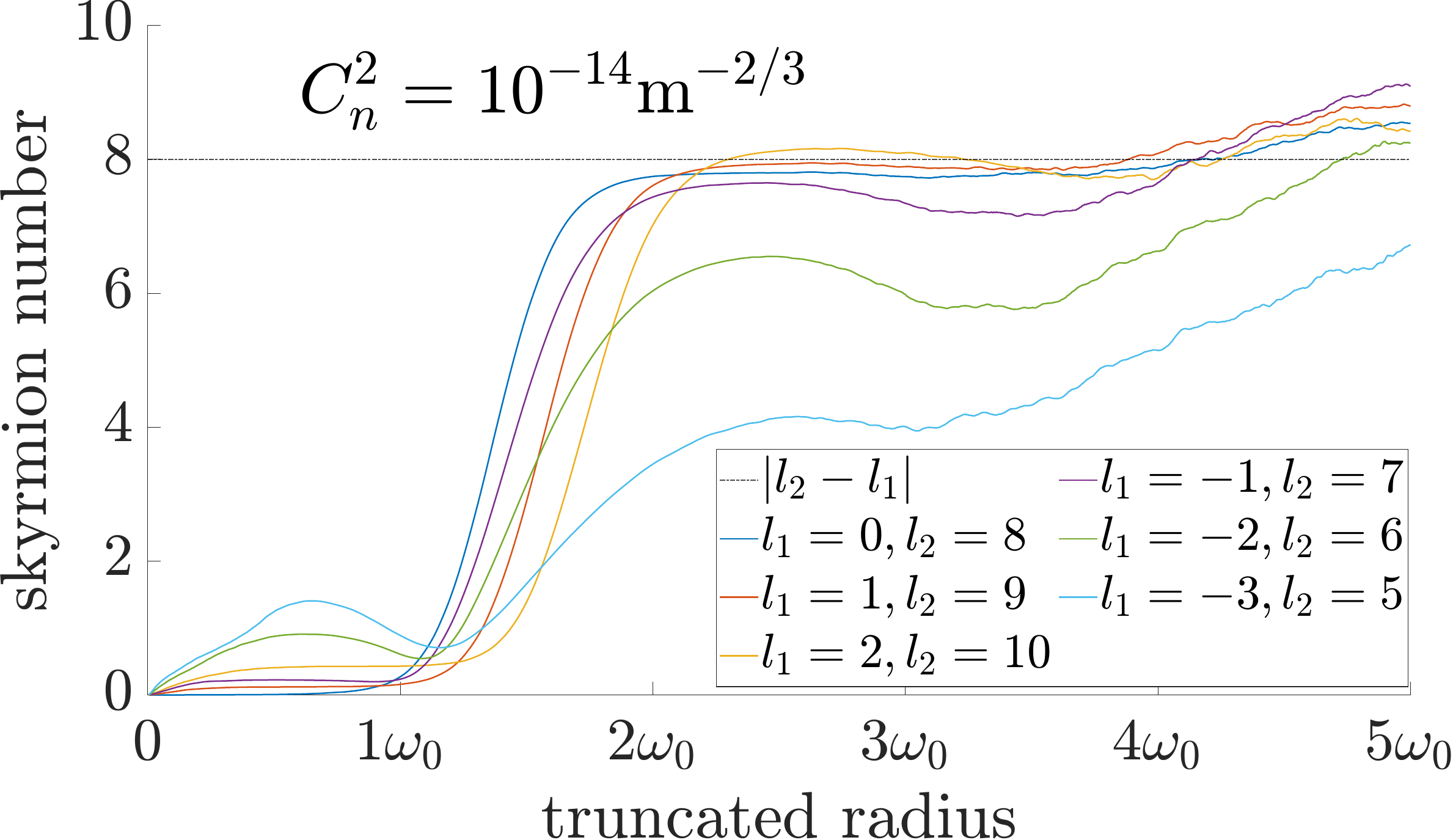}
        \label{fig:Nz_8_Cn2_-14}
    \end{minipage}
    }\subfigure[]{
    \begin{minipage}[t]{0.49\linewidth}
        \centering
        \includegraphics[width=2.2in]{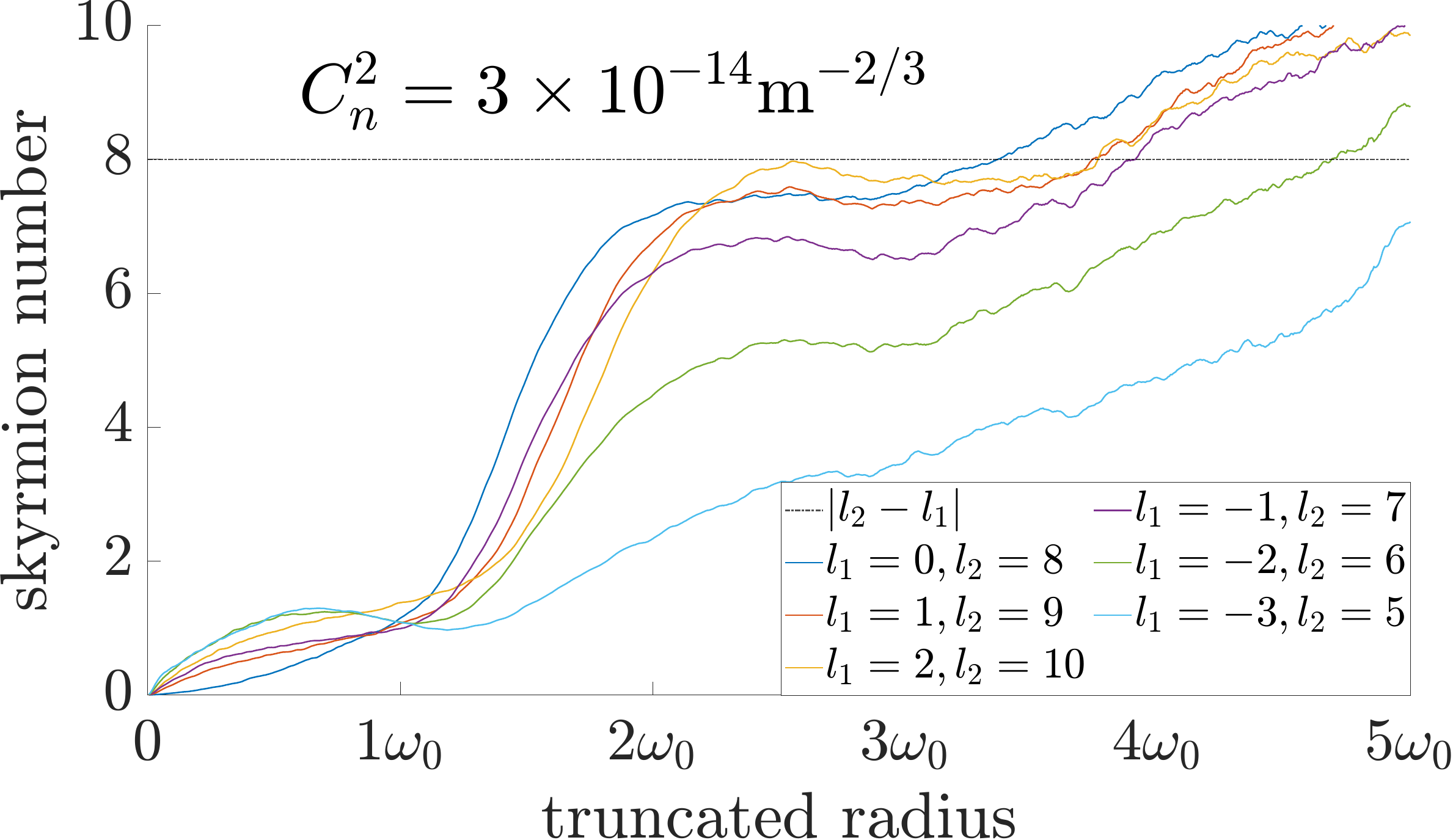}
        \label{fig:Nz_8_Cn2_3e-14}
    \end{minipage}
    }
    \caption{When fixing $\Delta l=8$, the numerical results $N_z$ vary with the truncated radius under weak to strong turbulence. There are six combinations of modes: $(0,8), (1,9), (2,10), (-1,7), (-2,6), (-3,5)$.}
    \label{fig:Nz=8}
\end{figure*}

Consequently, on the one hand, these numerical results in Fig. \ref{skyrmion_number_curve} suggest that skyrmion numbers of optical skyrmion fields have a certain degree of topological stability that is resistant to atmospheric turbulence. On the other hand, a larger difference in the absolute values of the two spatial mode numbers of the skyrmion field, i.e., $\Delta |l| = ||l_1|-|l_2||$, results in a reduced overlap of beam spots. This, in turn, contributes to a more robust preservation of the skyrmion number.\par
We fix the skyrmion number $N_z=8$ and vary the combinations of the two modes including $(0,8), (1,9), (2,10), (-1,7), (-2,6), (-3,5)$. Fig. \ref{fig:Nz=8} shows that skyrmion numbers for those six combinations vary with the truncation radius of the circle under different perturbations. In the stable region of the curve, skyrmion numbers obtained by the first five combinations are close to the theoretical value of $8$, but the result of the last combination $(-3,5)$ deviates from the value of $8$ when beams propagate in turbulence with $C_n^2 = \{10^{-16}\mathrm{m}^{-2/3}, 10^{-15}\mathrm{m}^{-2/3}\}$, as seen in Fig. \ref{fig:Nz_8_Cn2_-16}, \ref{fig:Nz_8_Cn2_-15}. In Fig. \ref{fig:Nz_8_Cn2_-14}, there is a large difference between the theoretical value and the numerical skyrmion numbers acquired by the last two combinations $(-2,6)$ and $(-3,5)$ with increasing turbulence strength ($C_n^2 = 10^{-14}\mathrm{m}^{-2/3}$). Moreover, when $C_n^2$ increases to $3\times 10^{-14}\mathrm{m}^{-2/3}$, the stability of the mode combination of $(-1,7)$ is also decreased in Fig. \ref{fig:Nz_8_Cn2_3e-14}. Thus, we can see that the larger the $\Delta |l|$, the larger resistant to the turbulent disturbance.

\section{Conclusion}
\label{sec:conclusion}
\noindent We employ a state-of-the-art model to simulate atmospheric turbulence characterized by a modified Kolmogorov power spectrum of the refractive index and a known split-step method that incorporates both free-space propagation and phase screen insertion. When a LG mode traverses through atmospheric turbulence, its intensity and phase patterns are distorted, and naturally, turbulence will cause deleterious effects in terms of information communication and channel transmission, especially if the LG mode is carrying valuable information. However, this paper focuses on a specific type of vector optical field known as the skyrmion field, which is composed of two polarization components with orthogonal spatial modes and possesses non-trivial topological properties that prevent a smooth transition between different skyrmion numbers \cite{ornelas2024non,hatcher2002algebraic}. We here assume that the light field is perfectly monochromatic, neglecting dispersion. However, it is important to note that when the light is not strictly monochromatic, dispersion becomes a significant factor, which could potentially lead to a decrease in coherence and alter the scenario. Nonetheless, in the context of the atmospheric environment, dispersion is deemed less significant when compared to phase disturbances.\par
Despite the fact that the intensity, phase, and even the polarization distributions of skyrmionic beams alter in turbulence, we primarily focus on the skyrmion number, a topological descriptor of the field’s characteristics. This skyrmion number, to some extent, is resilient to the turbulence present in atmospheric conditions. Additionally, our numerical demonstrations have also shown the skyrmion number becomes even more robust when the difference in the absolute values of $l_1$ and $l_2$ within each group is larger. The quantity has numerous practical applications such as communication, cryptography, and other related fields, and it holds great potential for further development. Furthermore, the more intricate the structure of the skyrmions, the more robust they become. Apart from the space-polarization skyrmions, space-time skyrmions pulses \cite{zdagkas2022observation,wang2023free,shen2024nondiffracting} and even 3D skyrmions like hopfions \cite{shen2023topological} are also considered as novel information carriers in recent years. Therefore, their properties during actual atmospheric propagation under turbulence require further investigation, which will have significant practical implications.

\appendix
\setcounter{table}{0}
\setcounter{figure}{0}
\setcounter{section}{0}
\setcounter{equation}{0}

\section{Additional details of selection of the truncated area}
\label{app:truncation area}
\noindent If we calculate the skyrmion numbers after propagating with and without turbulence, we should integrate them in space of two dimensions. Skyrmion numbers are dependent on the relative intensity proportion of two orthogonal components. Therefore, skyrmion numbers are greatly influenced by fluctuations in lower intensity regions and that is why we make a circular truncation. For consistency, the truncation radius is measured in units of the beam waist $w_0$. The screen size be $D = 0.5$m, the maximal truncation radius can reach $\frac{D}{2} = 5w_0$.\par
In fact, when the truncation radius is equal to the spot radius (the spot radius is the same as $w(z)$) plus the Gaussian spot radius (the spot radius of the fundamental mode $LG_{p=0,l=0}$), the entire spot can be thought to be completely included, and the excess may cause a sharp wobble in skyrmion numbers due to the weak intensity, just as seen in Fig.  \ref{fig:truncation_6} at the end of the curve warped. Thus we narrow the range of the integral domain and get a stable result as the desired skyrmion number. Here, the radius for each spatial beam can be calculated by the deviation $\Delta \boldsymbol{r}$ of $\boldsymbol{r}$ in the polar coordinate $(\boldsymbol{r},z)$, and the intensity distribution is taken as the probability distribution.\par
\begin{figure*}[htbp]
    \centering
    \subfigure[$\Delta l=1$]{
    \begin{minipage}[t]{0.33\linewidth}
        \centering
        \includegraphics[width=1.7in]{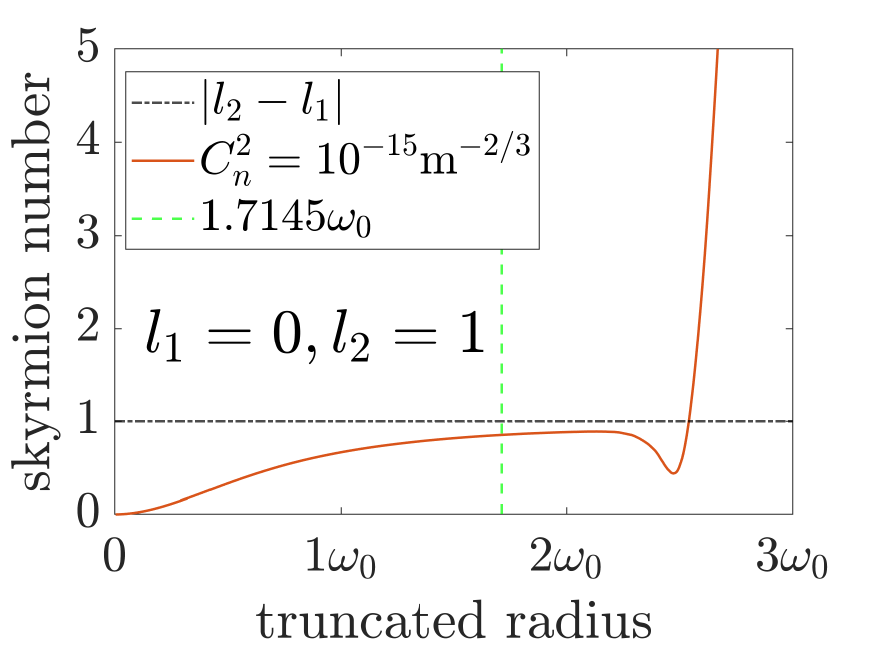}
        \label{fig:Appendix_L1_0_L2_1_Cn2_-15}
    \end{minipage}
    }\subfigure[$\Delta l=3$]{
    \begin{minipage}[t]{0.33\linewidth}
        \centering
        \includegraphics[width=1.6in]{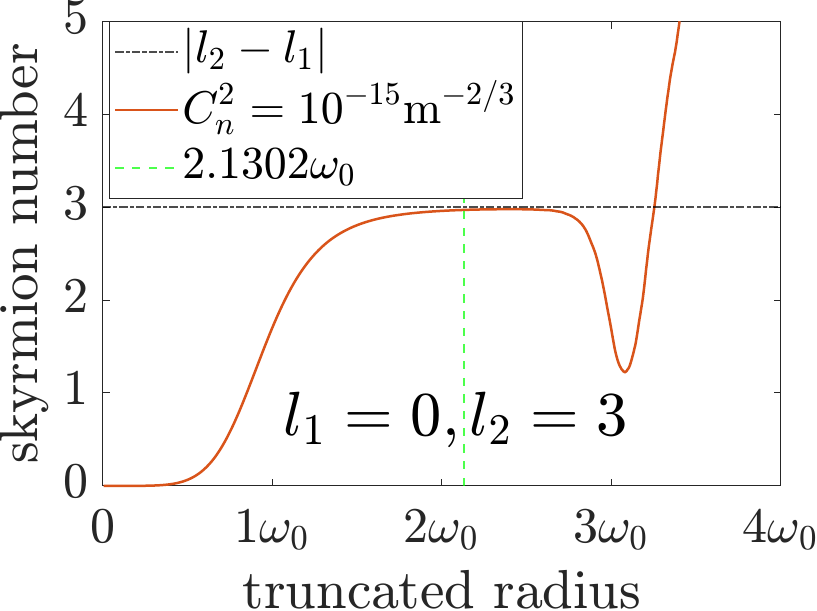}
        \label{fig:Appendix_L1_0_L2_3_Cn2_-15}
    \end{minipage}
    }\subfigure[$\Delta l=8$]{
    \begin{minipage}[t]{0.33\linewidth}
        \centering
        \includegraphics[width=1.7in]{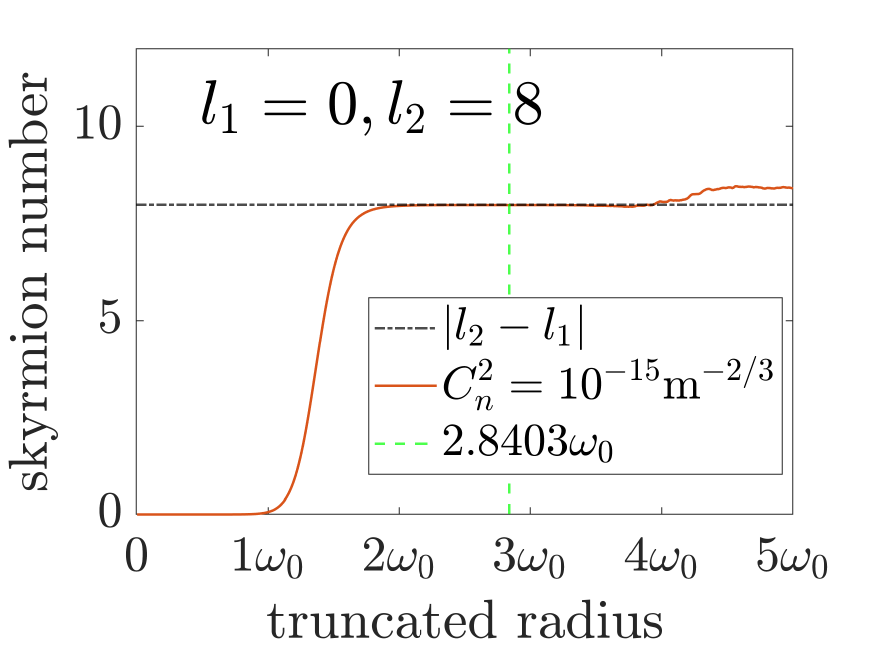}
        \label{fig:Appendix_L1_0_L2_8_Cn2_-15}
    \end{minipage}
    }

    \subfigure[$\Delta l=1$]{
    \begin{minipage}[t]{0.33\linewidth}
        \centering
        \includegraphics[width=1.7in]{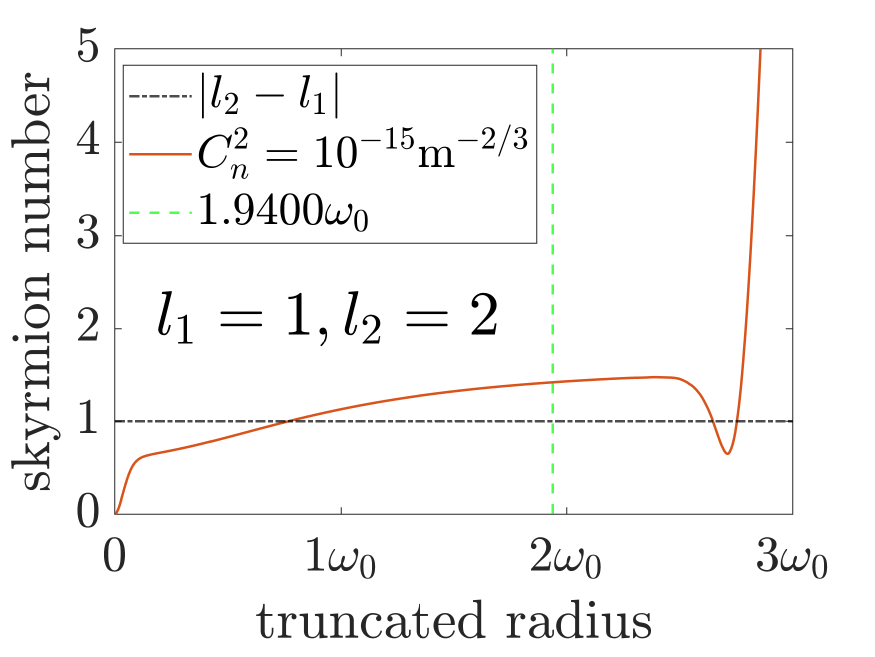}
        \label{fig:Appendix_L1_1_L2_2_Cn2_-15}
    \end{minipage}
    }\subfigure[$\Delta l=3$]{
    \begin{minipage}[t]{0.33\linewidth}
        \centering
        \includegraphics[width=1.6in]{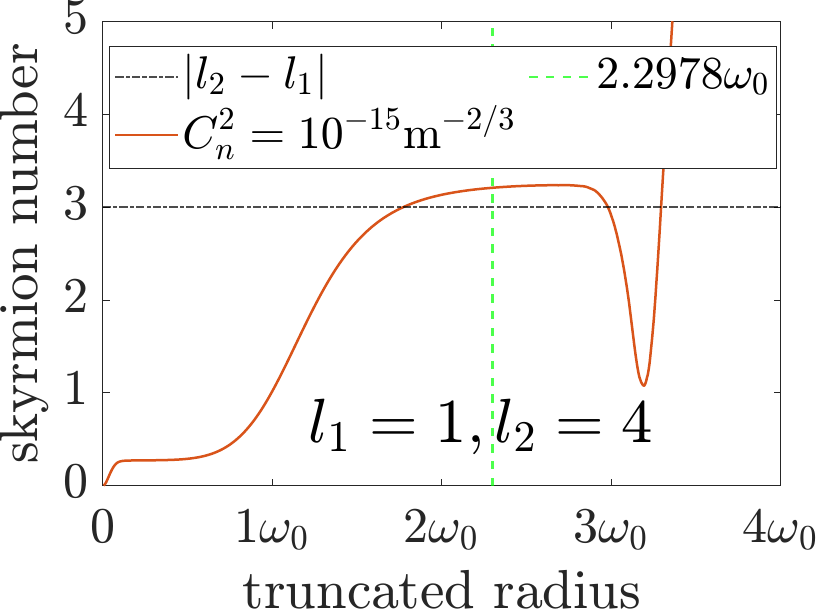}
        \label{fig:Appendix_L1_1_L2_4_Cn2_-15}
    \end{minipage}
    }\subfigure[$\Delta l=8$]{
    \begin{minipage}[t]{0.33\linewidth}
        \centering
        \includegraphics[width=1.7in]{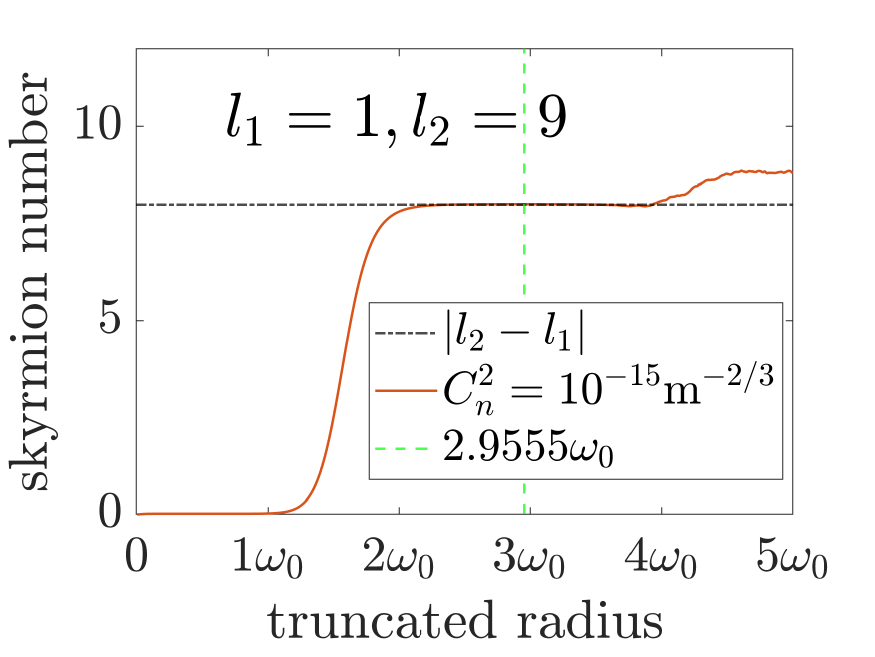}
        \label{fig:Appendix_L1_1_L2_9_Cn2_-15}
    \end{minipage}
    }

    \subfigure[$\Delta l=3$]{
    \begin{minipage}[t]{0.33\linewidth}
        \centering
        \includegraphics[width=1.55in]{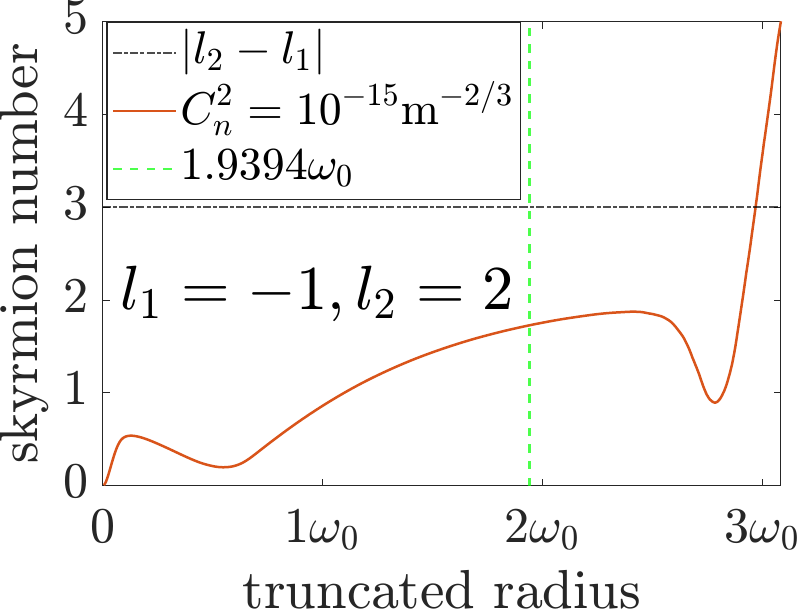}
        \label{fig:Appendix_L1_-1_L2_2_Cn2_-15}
    \end{minipage}
    }\subfigure[$\Delta l=8$]{
    \begin{minipage}[t]{0.33\linewidth}
        \centering
        \includegraphics[width=1.73in]{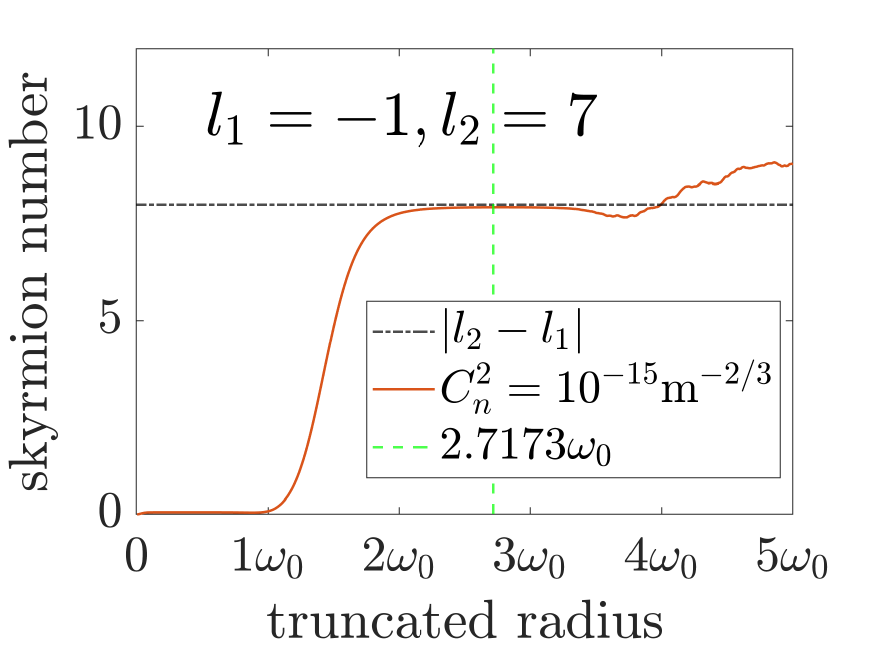}
        \label{fig:Appendix_L1_-1_L2_7_Cn2_-15}
    \end{minipage}
    }\subfigure[$\Delta l=8$]{
    \begin{minipage}[t]{0.33\linewidth}
        \centering
        \includegraphics[width=1.73in]{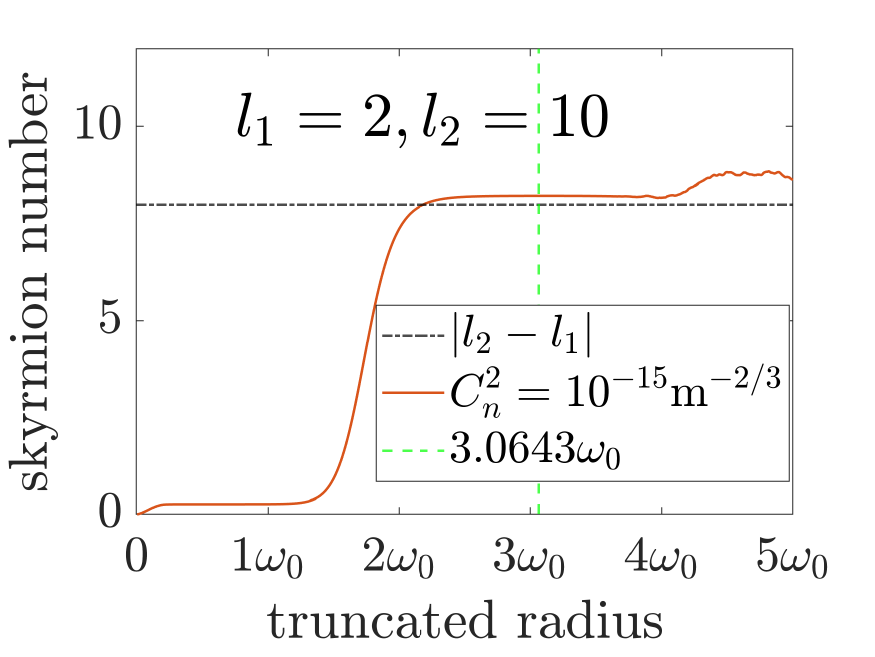}
        \label{fig:Appendix_L1_2_L2_10_Cn2_-15}
    \end{minipage}
    }
    \caption{Parameters: $C_n^2 = 1\times 10^{-15}\mathrm{m}^{-2/3}$. Different skyrmion numbers vary with the truncated radius. The green dashed vertical line in each subgraph shows that the truncation radius is the spot radius plus the Gaussian spot radius.}
    \label{fig:Appendix_L1_L2}
\end{figure*}
\begin{figure*}[t]
    \centering
    \subfigure[$l_1=0, l_2=1, \Delta l=1$]{
    \begin{minipage}[t]{0.33\linewidth}
        \centering
        \includegraphics[width=1.62in]{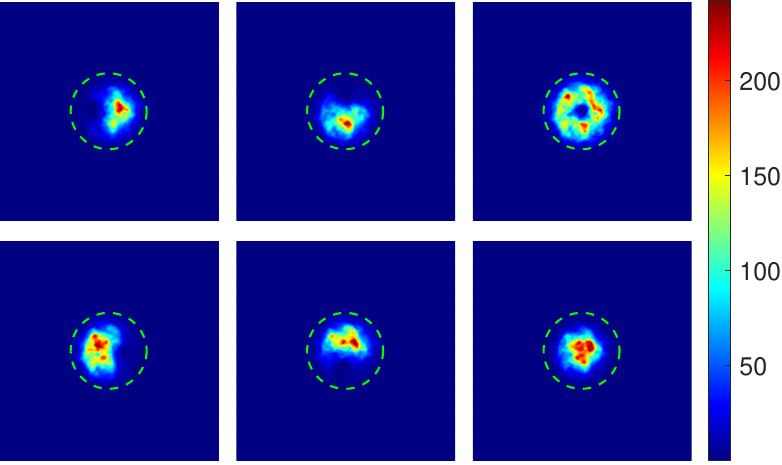}
        \label{fig:Appendix_basis_L1_0_L2_1_Cn2_-15}
    \end{minipage}
    }\subfigure[$l_1=0, l_2=3, \Delta l=3$]{
    \begin{minipage}[t]{0.33\linewidth}
        \centering
        \includegraphics[width=1.62in]{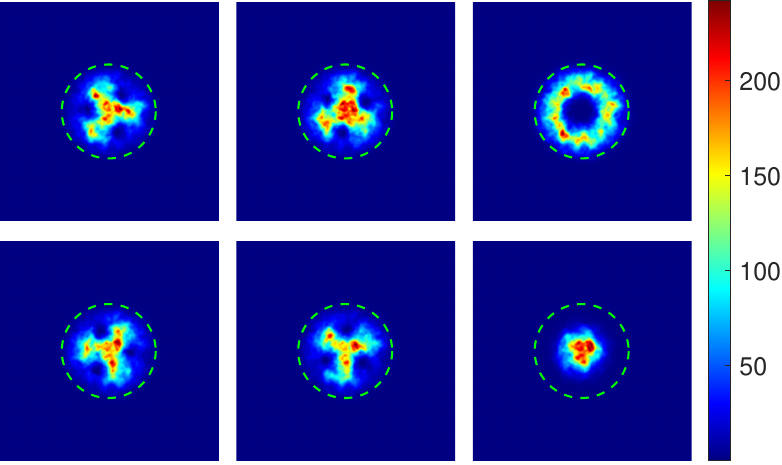}
        \label{fig:Appendix_basis_L1_0_L2_3_Cn2_-15}
    \end{minipage}
    }\subfigure[$l_1=0, l_2=8, \Delta l=8$]{
    \begin{minipage}[t]{0.33\linewidth}
        \centering
        \includegraphics[width=1.62in]{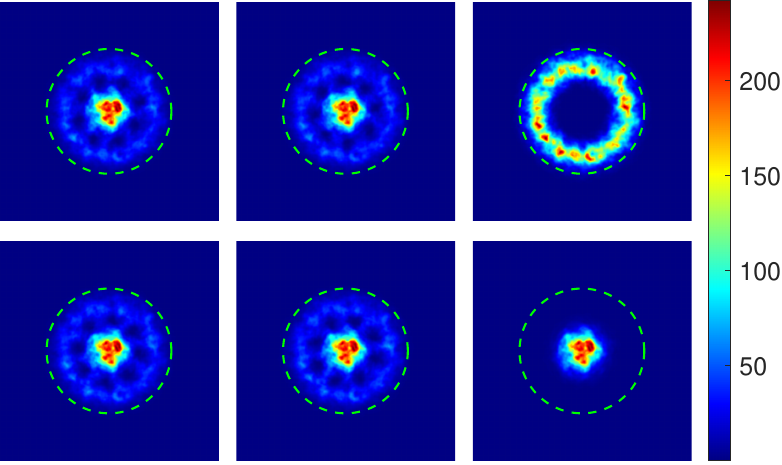}
        \label{fig:Appendix_basis_L1_0_L2_8_Cn2_-15}
    \end{minipage}
    }

    \subfigure[$l_1=1, l_2=2, \Delta l=1$]{
    \begin{minipage}[t]{0.33\linewidth}
        \centering
        \includegraphics[width=1.62in]{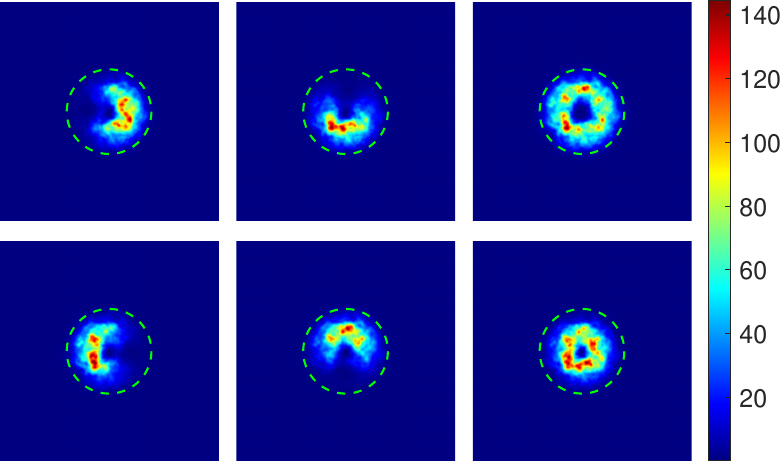}
        \label{fig:Appendix_basis_L1_1_L2_2_Cn2_-15}
    \end{minipage}
    }\subfigure[$l_1=1, l_2=4, \Delta l=3$]{
    \begin{minipage}[t]{0.33\linewidth}
        \centering
        \includegraphics[width=1.62in]{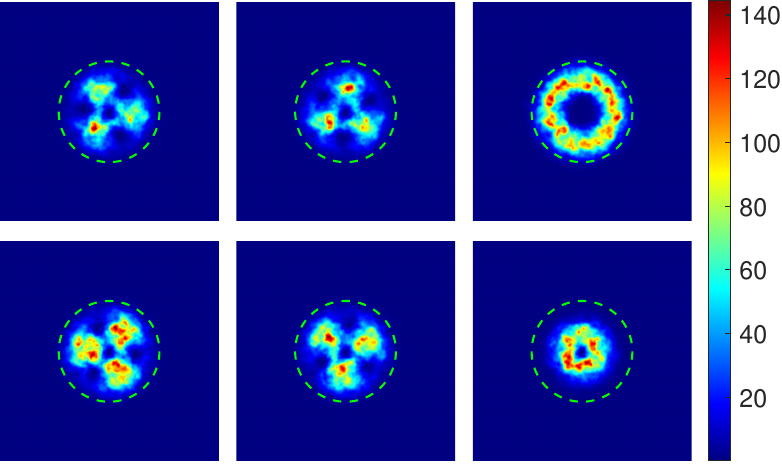}
        \label{fig:Appendix_basis_L1_1_L2_4_Cn2_-15}
    \end{minipage}
    }\subfigure[$l_1=1, l_2=9, \Delta l=8$]{
    \begin{minipage}[t]{0.33\linewidth}
        \centering
        \includegraphics[width=1.62in]{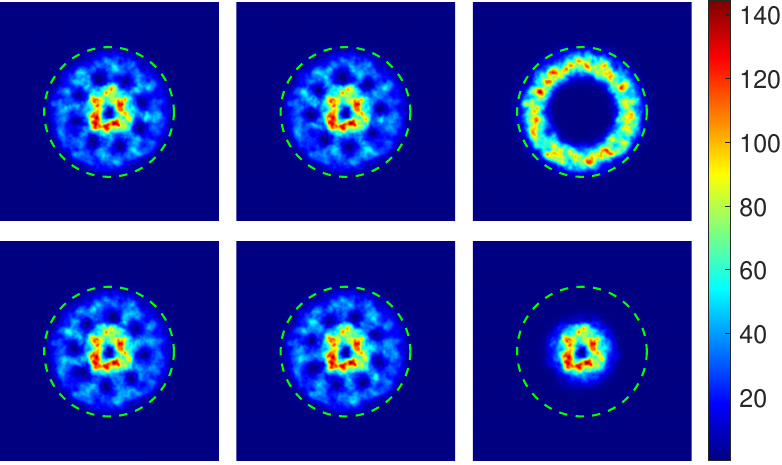}
        \label{fig:Appendix_basis_L1_1_L2_9_Cn2_-15}
    \end{minipage}
    }

    \subfigure[$l_1=-1, l_2=2, \Delta l=3$]{
    \begin{minipage}[t]{0.33\linewidth}
        \centering
        \includegraphics[width=1.62in]{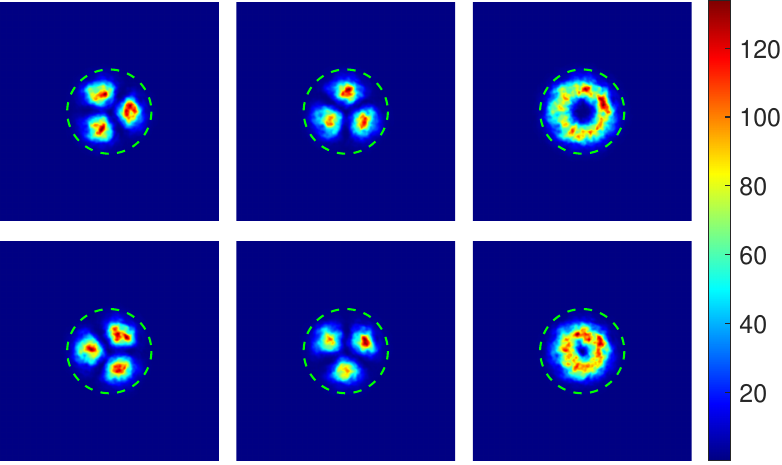}
        \label{fig:Appendix_basis_L1_-1_L2_2_Cn2_-15}
    \end{minipage}
    }\subfigure[$l_1=-1, l_2=7, \Delta l=8$]{
    \begin{minipage}[t]{0.33\linewidth}
        \centering
        \includegraphics[width=1.62in]{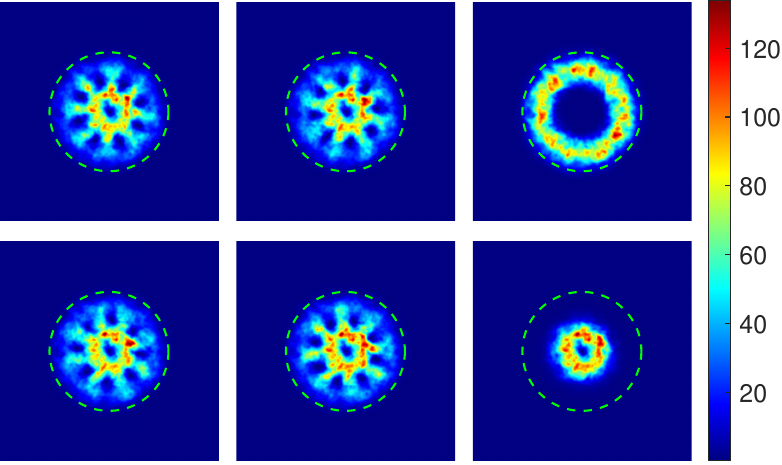}
        \label{fig:Appendix_basis_L1_-1_L2_7_Cn2_-15}
    \end{minipage}
    }\subfigure[$l_1=2, l_2=10, \Delta l=8$]{
    \begin{minipage}[t]{0.33\linewidth}
        \centering
        \includegraphics[width=1.62in]{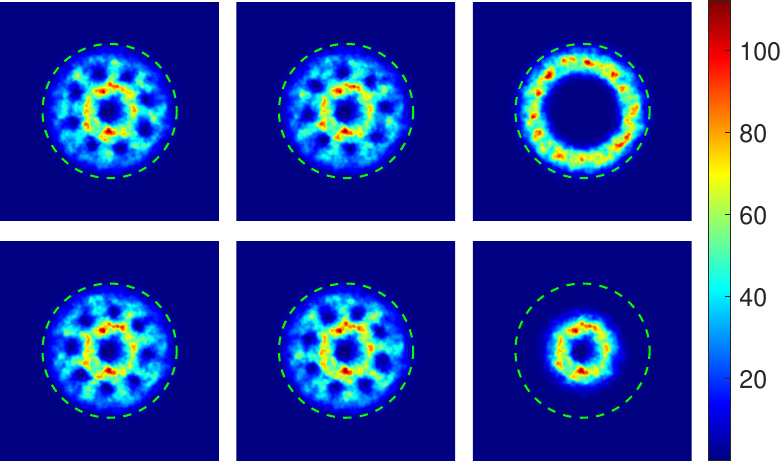}
        \label{fig:Appendix_basis_L1_2_L2_10_Cn2_-15}
    \end{minipage}
    }
    \caption{Parameters: $C_n^2 = 1\times 10^{-15}\mathrm{m}^{-2/3}$. The above subfigures correspond to the six eigenstates of the Pauli operator in Fig. \ref{fig:Appendix_L1_L2}, respectively.}
    \label{fig:Appendix_basis_L1_L2}
\end{figure*}

For instance, the results in Fig. \ref{fig:Appendix_L1_L2} and \ref{fig:Appendix_basis_L1_L2} indicate that it is appropriate to give a truncation with a definite truncated radius in the flat and stable area. The final skyrmion numbers are shown in Table \ref{tab:Appendix_Nz_Cn2_-15}, and these results require 1000 sampling averages due to the stochastic nature of turbulence.
\begin{table}[ht]
\setlength{\abovecaptionskip}{-0.5cm}
\setlength{\belowcaptionskip}{0.2cm}
\renewcommand{\arraystretch}{1}
    \centering
    \vspace{0.7em}
    \caption{The desired skyrmion numbers with the corresponding definite truncation radii for nine different mode combinations under medium turbulence. }
    \begin{tabular}{c|c}
    \toprule
    \diagbox{$(l_1, l_2)$}{$C_n^2(\mathrm{m}^{-2/3})$}& $1\times10^{-15}$ \\
         \midrule 
    $(0,1)$ & $0.8507\pm0.0046$\\
    $(0,3)$ & $2.9700\pm0.0032$\\ 
    $(0,8)$ & $7.9860\pm0.0005$\\
    $(1,2)$ & $1.4163\pm0.2433$\\
    $(1,4)$ & $3.2086\pm0.2037$\\
    $(1,9)$ & $8.0078\pm0.0301$\\
    $(-1,2)$ & $1.7285\pm0.2503$\\
    $(-1,7)$ & $7.9280\pm0.0726$\\ 
    $(2,10)$ & $8.2223\pm0.2354$\\
         \bottomrule
    \end{tabular}
    \label{tab:Appendix_Nz_Cn2_-15}
\end{table}

\section{Six projection bases of a skyrmion field with $l_1=0, l_2=8$}
\label{six proj bases with 0&8}
\begin{figure*}[htbp!]
    \centering
    \subfigure[$C_n^2 = 1\times 10^{-16}\mathrm{m}^{-2/3}$]{
    \begin{minipage}[t]{0.49\linewidth}
        \centering
        \includegraphics[width=1.8in]{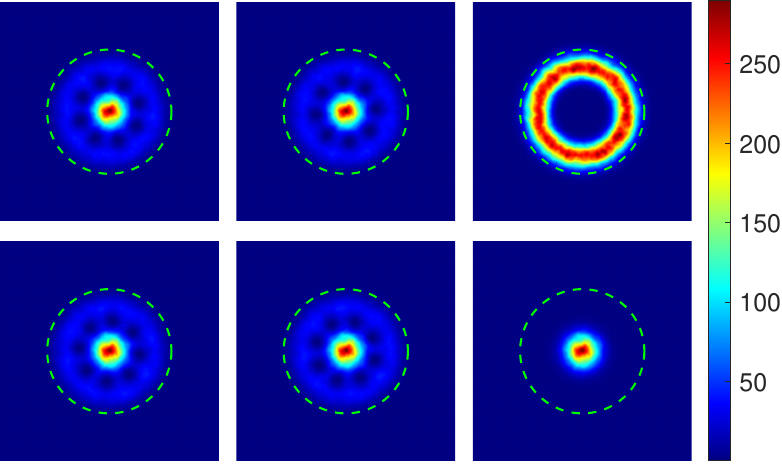}
        \label{app:basis_0_8_001}
    \end{minipage}
    }\subfigure[$C_n^2 = 1\times 10^{-15}\mathrm{m}^{-2/3}$]{
    \begin{minipage}[t]{0.49\linewidth}
        \centering
        \includegraphics[width=1.8in]{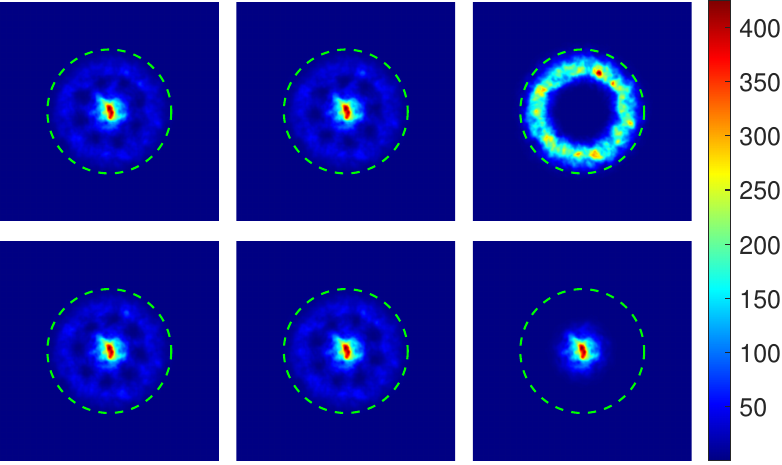}
        \label{app:basis_0_8_002}
    \end{minipage}
    }
    
    \subfigure[$C_n^2 = 1\times 10^{-14}\mathrm{m}^{-2/3}$]{
    \begin{minipage}[t]{0.49\linewidth}
        \centering
        \includegraphics[width=1.8in]{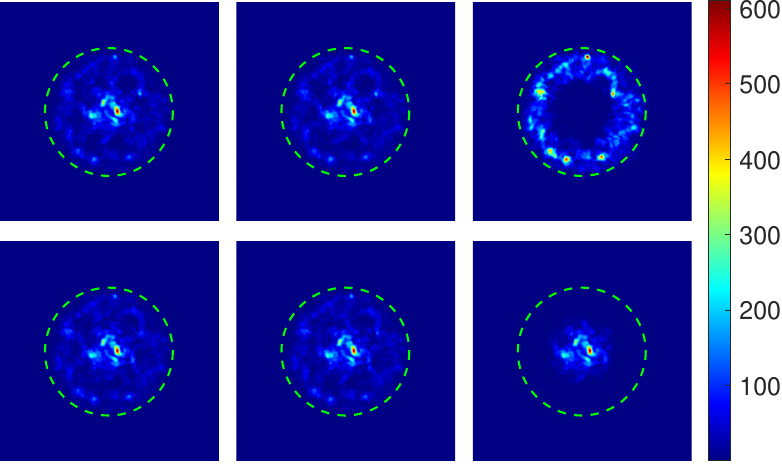}
        \label{app:basis_0_8_003}
    \end{minipage}
    }\subfigure[$C_n^2 = 3\times 10^{-14}\mathrm{m}^{-2/3}$]{
    \begin{minipage}[t]{0.49\linewidth}
        \centering
        \includegraphics[width=1.8in]{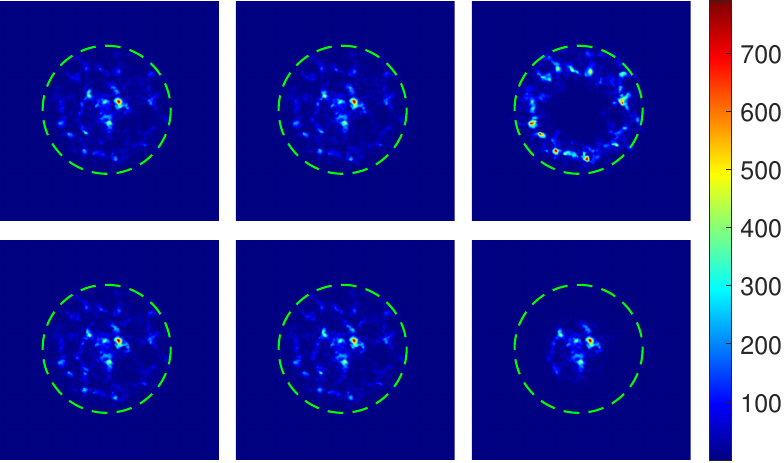}
        \label{app:basis_0_8_004}
    \end{minipage}
    }
    \caption{The intensity distribution of the field in the turbulence projected on six eigenstates of the Pauli operator, and the six eigenstates are $\{\ket{D}, \ket{A}, \ket{L}, \ket{R}, \ket{H}, \ket{V}\}$. Those results correspond to the theoretical skyrmion number of $8$ that the initial two spatial modes are $l_1=0$ and $l_2=8$, respectively. (a): $C_n^2 = 1\times 10^{-16}\mathrm{m}^{-2/3}$; (b): $C_n^2 = 1\times 10^{-15}\mathrm{m}^{-2/3}$; (c): $C_n^2 = 1\times 10^{-14}\mathrm{m}^{-2/3}$; (d): $C_n^2 = 3\times 10^{-14}\mathrm{m}^{-2/3}$. Green circles: the truncation regions for calculating more accurate skyrmion numbers.}
    \label{app-fig:basis_6_0&8}
\end{figure*}

In Fig. \ref{app-fig:basis_6_0&8}, each subgraph represents the numerically measured results of six bases for one parameter case. In Fig. B.\ref{app:basis_0_8_001}-B.\ref{app:basis_0_8_004}, $l_1 = 0, l_2 = 8$,  with increasing turbulence strength, the intensity distribution of each basis exhibits perturbation to a certain extent. The green circles indicate the selected truncation regions.\par

\section{Perturbations of skyrmion field with $l_1=1, l_2=9$}
\label{app:skyrmion field with l1=1&l2=9}
\begin{figure*}[htp]
    \centering
    \subfigure[$C_n^2 = 1\times 10^{-16}\mathrm{m}^{-2/3}$]{
    \begin{minipage}[t]{0.49\linewidth}
        \centering
        \includegraphics[width=1.8in]{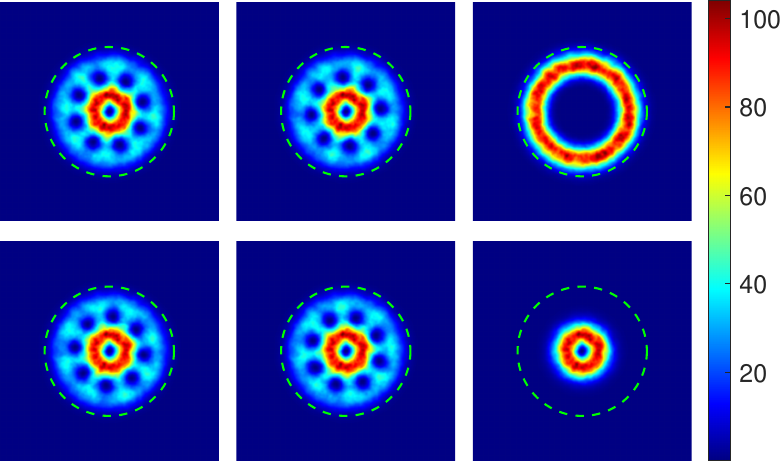}
        \label{app:basis_1_9_001}
    \end{minipage}
    }\subfigure[$C_n^2 = 1\times 10^{-15}\mathrm{m}^{-2/3}$]{
    \begin{minipage}[t]{0.49\linewidth}
        \centering
        \includegraphics[width=1.8in]{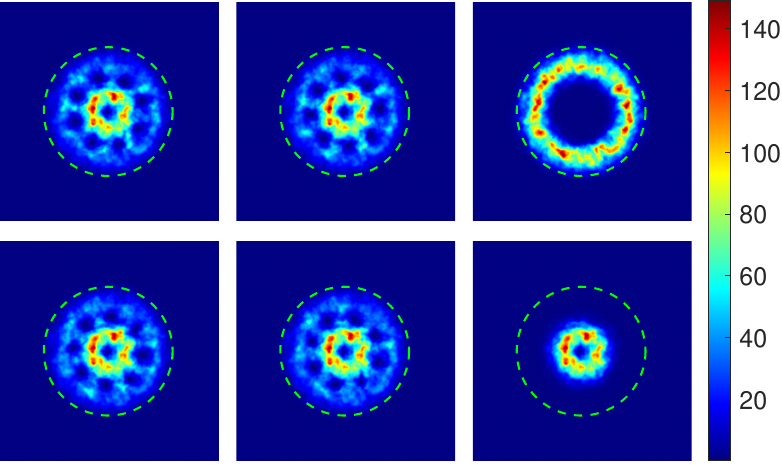}
        \label{app:basis_1_9_002}
    \end{minipage}
    }
    
    \subfigure[$C_n^2 = 1\times 10^{-14}\mathrm{m}^{-2/3}$]{
    \begin{minipage}[t]{0.49\linewidth}
        \centering
        \includegraphics[width=1.8in]{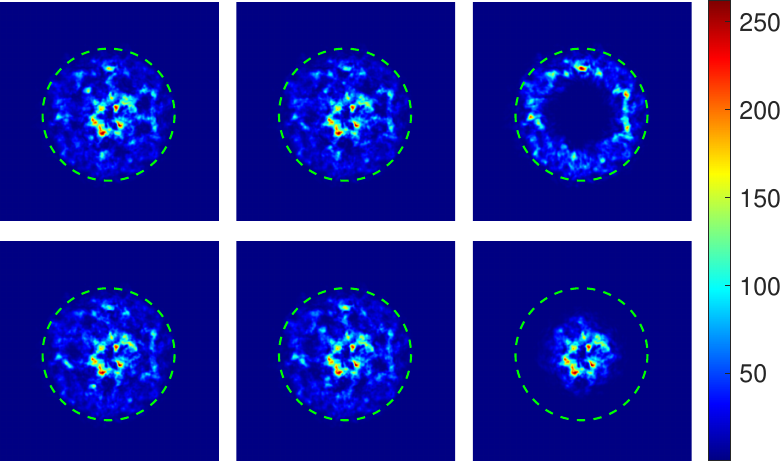}
        \label{app:basis_1_9_003}
    \end{minipage}
    }\subfigure[$C_n^2 = 3\times 10^{-14}\mathrm{m}^{-2/3}$]{
    \begin{minipage}[t]{0.49\linewidth}
        \centering
        \includegraphics[width=1.8in]{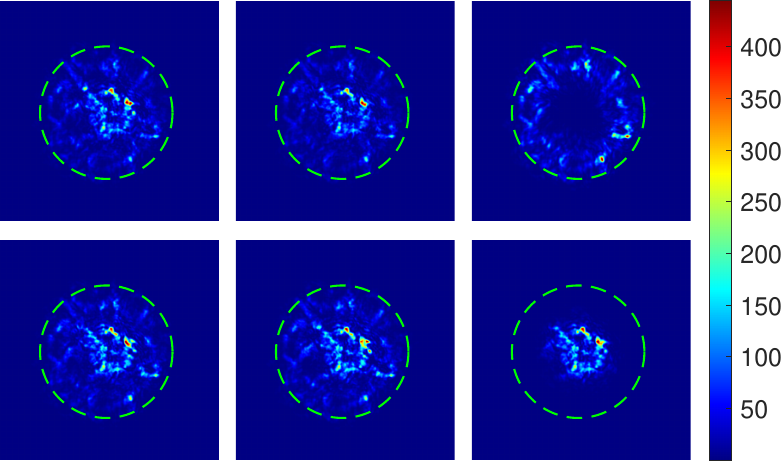}
        \label{cbasis_1_9_004}
    \end{minipage}
    }
    \caption{The intensity distribution of the field in the turbulence projected on six eigenstates of the Pauli operator, and the six eigenstates are $\{\ket{D}, \ket{A}, \ket{L}, \ket{R}, \ket{H}, \ket{V}\}$. Those results correspond to the theoretical skyrmion number of $8$ that the initial two spatial modes are $l_1=1$ and $l_2=9$, respectively. (a): $C_n^2 = 1\times 10^{-16}\mathrm{m}^{-2/3}$; (b): $C_n^2 = 1\times 10^{-15}\mathrm{m}^{-2/3}$; (c): $C_n^2 = 1\times 10^{-14}\mathrm{m}^{-2/3}$; (d): $C_n^2 = 3\times 10^{-14}\mathrm{m}^{-2/3}$. Green circles: the truncation regions for calculating more accurate skyrmion numbers.}
    \label{app-fig:basis_6_1&9}
\end{figure*}

In Fig. \ref{app-fig:basis_6_1&9}, it shows the intensity distribution of each basis of an optical skyrmion field propagating in atmospheric turbulence of different strengths and this field has initial spatial modes of $l_1=1, l_2=9$. 

\begin{figure*}[t]
    \centering
    \subfigure[$C_n^2 = 1\times 10^{-16}\mathrm{m}^{-2/3}$]{
    \begin{minipage}[t]{0.49\linewidth}
        \centering
        \includegraphics[width=1.8in]{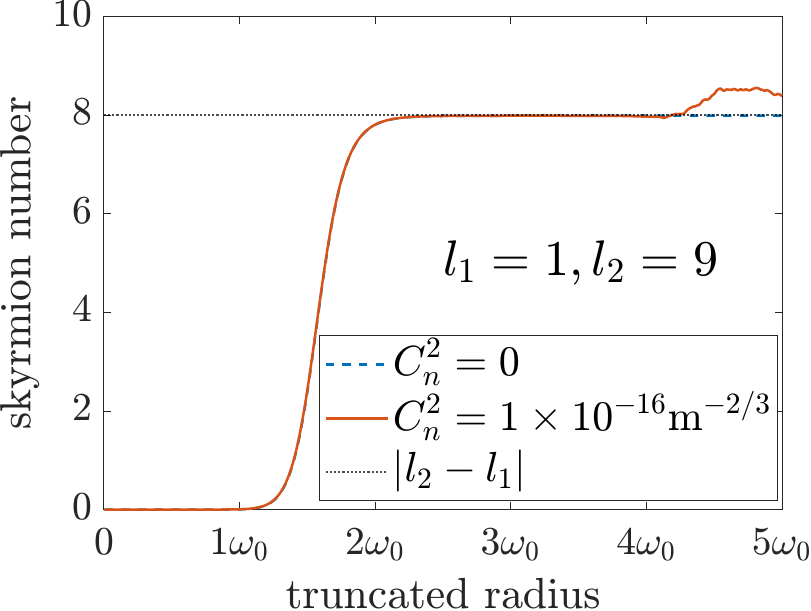}
        \label{app:L1_9_001}
    \end{minipage}
    }\subfigure[$C_n^2 = 1\times 10^{-15}\mathrm{m}^{-2/3}$]{
    \begin{minipage}[t]{0.49\linewidth}
        \centering
        \includegraphics[width=1.8in]{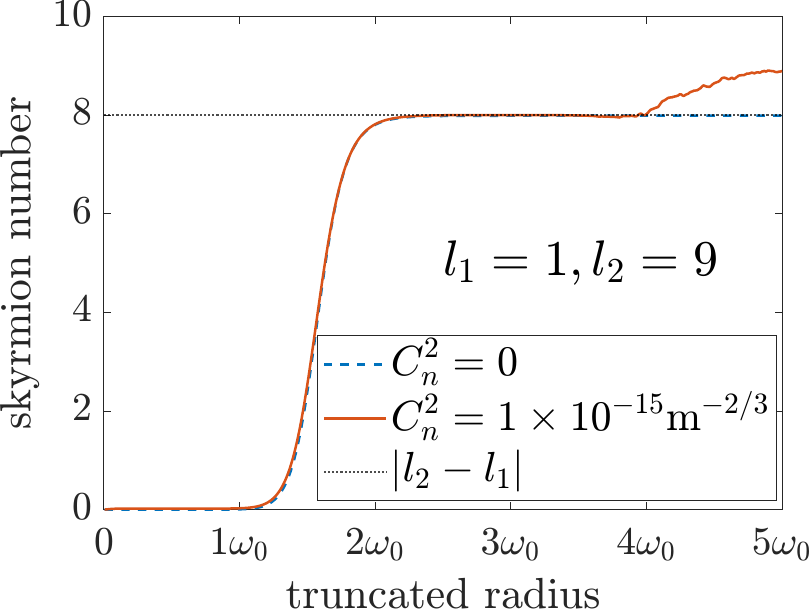}
        \label{app:L1_9_002}
    \end{minipage}
    }
    
    \subfigure[$C_n^2 = 1\times 10^{-14}\mathrm{m}^{-2/3}$]{
    \begin{minipage}[t]{0.49\linewidth}
        \centering
        \includegraphics[width=1.8in]{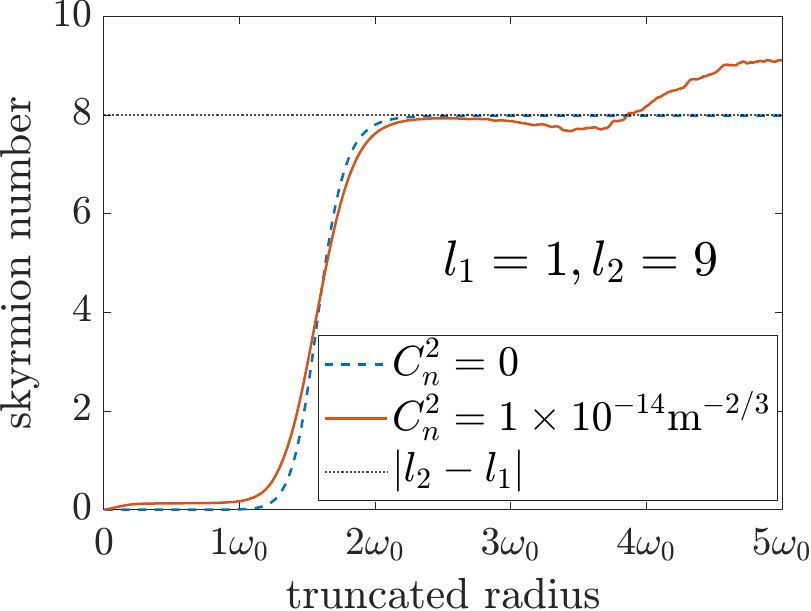}
        \label{app:L1_9_003}
    \end{minipage}
    }\subfigure[$C_n^2 = 3\times 10^{-14}\mathrm{m}^{-2/3}$]{
    \begin{minipage}[t]{0.49\linewidth}
        \centering
        \includegraphics[width=1.8in]{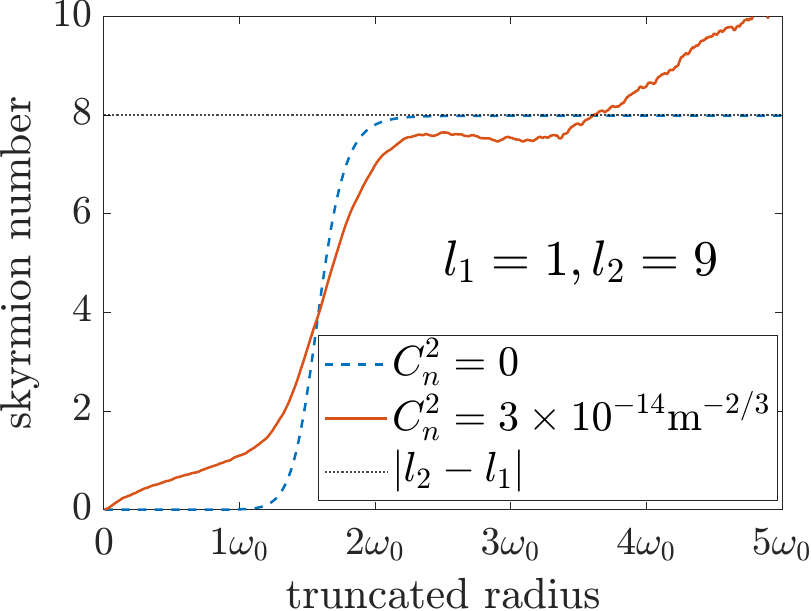}
        \label{app:L1_9_004}
    \end{minipage}
    }
    \caption{The curves of skyrmion numbers with the truncated radius. The black dotted line: the value $|l_2-l_1|$. The blue solid line: the skyrmion numbers in free space. The orange solid line: the skyrmion numbers in turbulence. (a)-(c): two OAM modes $l_1 = 1, l_2 = 9$, the turbulence refractive index constants $C_n^2 = \{1\times 10^{-16}\mathrm{m}^{-2/3}, 1\times 10^{-15}\mathrm{m}^{-2/3}, 1\times 10^{-14}\mathrm{m}^{-2/3}, 3\times 10^{-14}\mathrm{m}^{-2/3}\}$.}
    \label{app-fig:truncation_6_1&9}
\end{figure*}

And the corresponding curves of skyrmion numbers varying with truncated radius are displayed in Fig. \ref{app-fig:truncation_6_1&9}. The skyrmion numbers after the ensemble average are $7.9931 \pm 0.0011, 8.0078 \pm 0.0301, 7.8958 \pm 0.5890, 7.5663\pm 3.0678$, respectively. 
\vspace{3cm}

\section*{Declaration of competing interest}
The authors declare that they have no known competing financial interests or personal relationships that could have appeared to influence the work reported in this paper.

\section*{Data availability}
Data will be made available on request.

\section*{Acknowledgments}
L.-W.W. acknowledges and thanks Jie Zhu for valuable discussions and encouragement. This work is supported by the National Natural Science Foundation of China (Grant No. 92065113), the Innovation Program for Quantum Science and Technology (2021ZD0301201), the University Synergy Innovation Program of Anhui Province (Grant No. GXXT-2022-039).

\bibliographystyle{elsarticle-num} 
\bibliography{elsarticle-template-num.bbl}

\end{document}